\documentclass[12pt,fleqn]{article}
\usepackage[headings]{espcrc1}
\readRCS
$Id: espcrc2.tex,v 1.2 2004/02/24 11:22:11 spepping Exp $
\usepackage{epsfig}
\usepackage{amsmath}
\usepackage{amssymb}
\usepackage{euscript}
\usepackage{color}

\title{{ Luminosity measurement method for the LHC: 
{The detector requirements studies}}\thanks{This work was supported in part by the programme of
co-operation between the IN2P3 and Polish Laboratories No. 05-117, 
Polonium Programme No. 17783NY and Polish Grant No. 665/N-CERN-ATLAS/2010/0.
}}
\author{ M.~W. Krasny\address[LPNHE]{ LPNHE, Pierre and Marie Curie 
                         University, CNRS-IN2P3,  Tour 33, RdC, \\
                         4, pl. Jussieu, 75005 Paris, France.}, %
J. Chwastowski\address[PK]{ Institute of Teleinformatics, 
Faculty of Physics, Mathematics and Computer Science,
                                    Cracow University of Technology,
                ul. Warszawska 24, 31-115 Krak\'ow, Poland.}
\address[IFJ]{ Institute of Nuclear Physics PAN,
ul. Radzikowskiego 152,
31-342 Krak\'ow, Poland.}, %
A. Cyz\addressmark[IFJ] %
and
K. S{\l}owikowski\addressmark[IFJ]}

\begin{document}

\maketitle

\vspace*{10mm} 

\begin{abstract}

Absolute normalisation of the LHC measurements with a precision of 
$\cal{O}$(1\%) is desirable  but beyond the reach of the present LHC detectors. 
This series of papers proposes and evaluates a measurement method capable to 
achieve such a precision target. In our earlier paper  \cite{first} we have selected the  
phase-space region where the lepton pair production cross section in $pp$
collisions at the LHC can be controlled with $\leq 1\%$ precision and  
is large enough to reach a comparable  statistical accuracy  of the 
absolute luminosity measurement on the day-by-day basis. 
In the present one  the performance  requirements for a dedicated detector, 
indispensable to efficiently select events in the  proposed phase-space region, are discussed. 
\end{abstract}

\vspace*{10mm} 

\section{Introduction}
\label{sec:Introduction}

For the direct searches for new particles  at the LHC the precision of the 
absolute normalisation of the measured cross sections is of 
secondary importance. However, to fully exploit the LHC discovery 
potential, direct searches must be complemented by precise
measurements of the Standard Model cross sections and, subsequently, by the scrutiny
of their compatibility with those determined at the previous colliders. 
Achieving the highest possible  precision of the absolute scale of the LHC cross sections 
is of utmost importance for such a complementary programme. 

Presently,  no viable measurement scheme capable to reach the  desired 
$\leq 1\%$ precision of the absolute scale of the cross sections exists. 
Moreover,  no scheme exists allowing  to cross-normalise the LHC cross 
sections measured at variable centre-of-mass energies and variable LHC beam particle species (ions)
to a desirable precision level of $\leq 0.1 \%$. 

This paper is the second one of the series of papers presenting a measurement method 
which is capable of achieving  the above precision goals. This method is based on 
the measurement of the rate of the opposite-charge lepton pairs produced in  peripheral 
collisions of the beams' particles.

In \cite{first} we have selected the kinematic region in which 
the  rate is large enough to reach a 1\% statistical precision of the luminosity
measurement on the day-by-day basis. We have demonstrated  that better than 1\% 
accuracy of the theoretical control of the pair rate can be reached by a 
restriction of their allowed phase-space to the region of small 
transverse momentum  leptons produced back-to-back
in the plane transverse to the collision axis. Such a 
restriction allowed us to  to suppress drastically  the contribution of 
the inelastic  collisions and those of elastic collisions in which the internal charge 
structure of protons is resolved.

Efficient selection of  electromagnetically produced lepton pairs in the proposed phase-space 
region,  out of a huge background of any unlike charge particle pairs produced 
in the ordinary, minimum bias collisions of the LHC beam particles,  represents a major challenge. 
For the  bunched beams,  colliding at large luminosity the challenge is twofold. 
The overall rejection power 
of the background hadron pairs of the order of $ \sim10^{10}$  must be achieved. 
However, the most difficult challenge is to drastically reduce the GHz-range rate of the
background pairs at the very early stage of the data selection process -- 
preferentially at the first trigger level  (LVL1). 
This is obviously  beyond the reach of the present general purpose LHC detectors.

Our  most important constraint for designing a luminosity detector which 
allows to meet the above requirements is that it should be fully incorporated within the fiducial volume 
of one of the already existing LHC detectors. In particular, its trigger and the 
data acquisition (TDAQ) system should  be a partition  of the host detector TDAQ. 
The rationale behind such a constraint  is twofold. 
Firstly, the host detector signals will play an important role in  the overall   
background rejection scheme, both for  the  high level trigger (HLT) 
and the off-line selection stages. Secondly,  
the host-detector offline reconstructed objects could be used  for  
precise monitoring of the luminosity detector performance.
The luminosity events, once accepted by the luminosity 
detector LVL1  trigger,  could  be exposed to  the host detector specific
event selection procedures in the same way as  all the other events. Owing to such a 
scheme, the dead time corrected luminosity can be directly measured\footnote{ 
Only in such a case the precision of the measured rate of the lepton pairs 
could be directly reflected in the  precision of the absolute normalisation of 
any event sample, regardless  of the time variation of:  the bunch-by-bunch 
beam intensity, the time dependent detector operation efficiency, and the TDAQ 
dead time.}.

For the studies of the luminosity detector performance requirements, presented in this paper,  
we have chosen ATLAS  \cite{ATLAS} as the host detector. This detector, following the 
decision of the ATLAS collaboration to stage the production 
of the TRT C-wheels, disposes an empty space which could, in 
principle, accommodate the luminosity detector.  This leftover  space 
happens to be the most optimal one for a detector satisfying all the requirements 
discussed  in this paper\footnote{The dead material budget of such a detector 
could be smaller than that of the TRT C-wheels and  its presence would not 
deteriorate the performance quality of the ATLAS detector.}.  

The paper is organised as follows. In section \ref{sec:strategy} the
general aspects of the event selection strategy are presented. The tools
and the simulation methods are discussed in section \ref{sec:tools}. In section
\ref{sec:geom} the luminosity detector geometry and the magnetic field
configurations are introduced. The characteristics of the background and the
signal events are presented in section \ref{sec:charac}. Section
\ref{sec:detperf} is devoted to the detector resolution requirements
discussion. The dead material effects are evaluated in section
\ref{sec:dead}. Finally, Êsection \ref{sec:TDAQ} is focussed on the
trigger and the data acquisition system requirements.

\section{An initial overview of the event selection strategy}
\label{sec:strategy}

In our  previous paper  \cite{first},  an optimal  phase-space region of  the lepton pair production process
$pp \rightarrow l^+l^- + X$,
 for the LHC  luminosity measurement was selected. 
Our choice was driven by the following two requirements: 
\begin{itemize}
\item  to assure a high rate  of the corresponding "luminosity" events,  
\item  to guarantee a   high precision of its theoretical control,   
\end{itemize}
while taking into account the real constraints
of the existing general-purpose LHC experiments. 
The optimal compromise was achieved by requesting 
leptons to be produced in the central 
pseudorapidity region, $-2.7 \leq \eta \leq 2.7$.  Their transverse 
momenta were requested to satisfy the conditions:  $p_T ^{l^+},p_T ^{l^-}  \geq 0.1$ GeV/c.  

The  ATLAS  detector  has the requisite capacity 
to measure both the tracks and the calorimetric energy deposits of such leptons.  However, in its present 
configuration it is unable to select on-line a requisite  fraction of events with leptons 
in the selected  phase-space region. 
Such a task is difficult because  lepton pairs,  produced in the selected kinematic region,  represent a $10^{-9}$ 
fraction of all the unlike charge hadron pairs produced in ordinary minimum-bias hadron processes.  

The detector upgrade requirements are thus driven by the  necessity of adding  the  triggering capacity for the 
luminosity events. Variety of  triggering strategies could be employed. Each of them would lead to a different 
requirements for the detector upgrade. The selection strategy advocated  below minimises  the 
interference with the host  detector layout and its  TDAQ architecture. More importantly, it minimises the usage  of 
the host detector TDAQ resources to a level which will hardy be noticeable.  

The ATLAS LVL1 trigger can accept events with maximum frequency of 100
kHz \cite{ATLASTDAQ}.  It is assumed that of the order of 1\% of this bandwidth 
could be assigned to the lepton pair production candidates. In such a case the initial rate of the charged 
hadron pairs has to  be reduced by a factor of $10^7$
already by the LVL1 trigger. The above rate reduction has to be based, almost exclusively,  
on the trigger signals coming from the dedicated luminosity detector.  
Once the initial rate of the lepton pair candidate events is reduced to a  ~1 kHz level a further selection
of events by the level 2 trigger (LVL2) and by  the event filter (EF), can be based entirely 
on the ATLAS detector signals. As a consequence the performance requirements for a dedicated 
luminosity detector will be specified, almost exclusively,  in terms of its  LVL1 triggering capacity. 

Several event-selection strategies based on the pair-by-pair rejection principle  were investigated. 
None of them was able to  satisfy the LVL1 latency requirement. Therefore,  compromises had  to be made. 

A  strategy capable of  achieving such a rejection factor must drastically 
reduce the rate of the bunch crossings which will be fully analysed by the luminosity detector LVL1 trigger logic. 
Moreover, the analysed  bunch crossings must have a significantly smaller multiplicity of charged particles
traversing the fiducial volume of the luminosity detector than the rejected bunch crossings.

In  \cite{first} we proposed to select  only the 
``silent bunch crossings" for the luminosity determination. 
 A ``silent bunch crossing'' was defined as a bunch crossing in which
none of the protons of the colliding  bunches had a strong interaction mediated collision. 
In this paper, the above definition will be reformulated using the using solely the LVL1 signals of 
the luminosity detector\footnote{In such a scheme, the instantaneous luminosity dependent 
fraction of the silent bunch crossings must be precisely monitored using a sample of random bunch crossing trigger.}.

The necessity of the  initial  reduction of  the rate of the fully analysed  bunch 
crossings excludes a direct applicability of the proposed method over the 
periods of the machine operation in which the delivered luminosity will be 
higher than $L\sim 3 \times 10^{33}$~cm$^{-2}$~s$^{-1}$ (if the luminosity is 
distributed uniformly among the colliding bunches).
The luminosity collected in such  machine operation periods will be 
determined by  a  complementary method\footnote{This method uses the 
minimum bias events to extrapolate  the measurement of the absolute 
luminosity to the "high-luminosity" periods.}.
 
 An important singularity of our event selection strategy is the choice of
events,  containing a pair of particles with highly collinear, back-to-back transverse momenta,  
already by the luminosity detector LVL1 trigger logic. Lepton identification is left to the subsequent
trigger levels. The rationale 
behind such a choice is that only the rate of such a subsample of events can be controlled 
theoretically with satisfactory precision \cite{first}.
The LVL1 selected sample is dominated by the charged hadron pairs 
produced in the ordinary minimum bias processes. 
This background can be determined experimentally to a very high precision
because minimum bias events are recorded parasitically in all the phases of the detector 
operation,  providing the requisite background monitoring data samples. 
The background subtraction scheme will thus be based entirely on the collected data 
and will be independent of all the modelling aspects of the soft hadron interactions. 

Another important aspect of the LVL1 trigger selection strategy presented in 
this paper is that it will be  based only on the topological properties of the events.  
Our general guiding principle 
was to try to express  the gauge invariance of electromagnetic interactions -- 
which determines  the basic properties of the signal events -- in terms of 
topological variables. Such variables  assure  a  robust event selection, fairly 
insensitive to the time-dependent aspects of the detector and machine operation.

\section{Tools and analysis methods}
\label{sec:tools}

The sample of the lepton-pair signal events was generated with the LPAIR \cite{LPAIR}
generator. This generator was upgraded to suit our needs (see \cite{first} 
for details). 

For the simulations of the minimum bias events the PYTHIA \cite{PYTHIA} event 
generator was chosen. The adequacy of this generator in describing the 
minimum bias events at the LHC is of secondary importance, for studies
presented below. 
All the efficiencies and acceptances of presented method will
be determined directly from the data using large statistics monitoring data 
samples recorded parasitically over the standard data taking  
periods. The PYTHIA generator is thus merely used to illustrate the event 
selection strategy and for the initial specification of the detector and the
trigger performance requirements.  An underestimation of the background level by
a factor of 10 will make the measurement more difficult but will,  by no means,  
invalidate the proposed measurement procedure.  

A classical method of optimisation of the detector and  the trigger designs
should ideally be based on the GEANT \cite{Geant} simulations of the 
signal and background events  in the fully specified detector. Such a method is, however, of little 
use for studies requiring  a  sample of $~10^{11}$ simulated background events.  
For such a large sample the events, we had to base our studies on  simplified 
methods of tracking of particles in the magnetic field, parametrised simulation
of their multiple scattering in the dead material,  and  
an approximate description of the effects of the photon radiation by electrons.

The Coulomb multiple scattering was simulated using a Gaussian approximation \cite{PDG}.
The photon bremsstrahlung  in the material was simulated using the 
Tsai  formulae \cite{tsai}. These simulation simplifications are justified by our aim to 
determine a safe upper limit of the rate of the background 
events and the lower limit of the rate of the luminosity signal 
events. For such a goal the developed tools are both adequate and sufficiently 
precise. 

\section{Detector geometry}
\label{sec:geom}
\begin{figure}[h]
\begin{center}
\epsfig{file=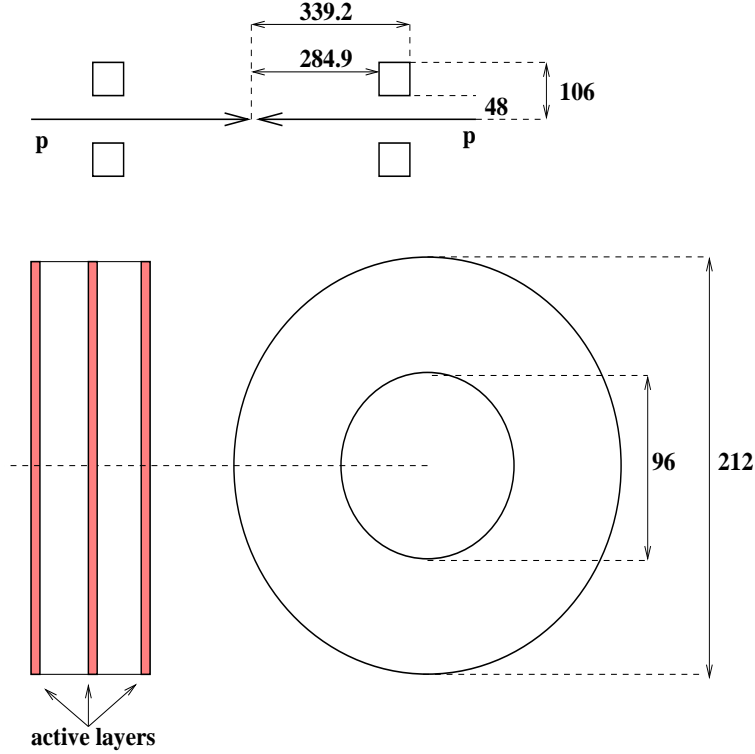,width=10cm,height=10cm}
\end{center}
\caption{The luminosity detector fiducial volume. Distances are given in centimetres}
\label{plot1}
\end{figure}

The geometry of the proposed  luminosity detector is shown in Fig. \ref{plot1}. The detector 
fiducial volume consist of two identical cylinders placed symmetrically with 
respect to the interaction point. The cylinder has the inner radius of 48 cm, 
the outer radius of 106 cm and the length 54.3 cm extending from $z_f = 284.9$ 
cm to $z_r = 339.2$ cm along the beam line.  Each cylinder is assumed to have 
three active layers (the $z_1, z_2, z_3$ planes) delivering the measurement of 
the position of a hit left by a particle. These planes are positioned 
at the distances of $z_1 = 285.8$, $z_2 = 312.05$ and $z_3 = 338.3$ cm away from 
the interaction point. 

The momentum spectrum and the multiplicity of the charged particles traversing the 
above defined fiducial volume depend upon  the magnetic field in which the 
charged particles propagate. In the studies presented in this paper  two field 
strengths  were considered: $B = 0$ T and $B = 2$ T of a uniform, solenoidal
magnetic field, labelled respectively  as the ${\bf B0}$ and ${\bf B2}$ 
configurations\footnote{These two configurations correspond to the  nominal  
and the "zero-current" configuration of the ATLAS central tracker solenoid.}.

A charged particle entering the fiducial volume of the luminosity detector by crossing the 
$z_1$ plane and leaving it by crossing the $z_3$ plane will be 
called hereafter the ``tagged particle''. Its hits in the detector planes 
define the luminosity detector ``track segment''. The lepton pair production candidate events 
are  identified by the presence of the two tagged particles 
in the $z > 0$ part of the detector volume and no tagged particles in the $z 
< 0$ part,  or equivalently by a  mirror reflection of such a configuration.
They will be denoted in the following  as the ``2+0'' events. 
Both,  the electromagnetic and the strong interactions of the beam particles 
may produce ``2+0'' event signatures. 

\section{Signal and background events}
\label{sec:charac}

The strong interaction induced collisions of protons, in particular the 
diffractive ones, contribute overwhelmingly  to the ``2+0'' event 
sample. Both, the rates of silent bunch crossings and that of the ``2+0'' events 
depend upon the bunch-bunch collision luminosity. For the  canonical operation of 
the LHC machine \cite{LHC} with 25 ns bunch spacing 
and at the highest achievable luminosities several proton-proton 
collisions may take place within one bunch crossing. It was found,  using the 
PYTHIA simulated events,  that the probability of two or more proton-proton collisions 
giving  rise to a  ``2+0'' signature is 0.02.   Consequently, 98\% of 
bunch crossings with the ``2+0'' signature have precisely one proton-proton 
collision. This observation allows us to simplify the background studies 
by considering the single proton-proton collision events as 
the dominant background source. All bunch crossings with two or more 
hadron interaction will be neglected in the estimation of the hadron 
background to the ``2+0'' event sample. As a consequence, in the following,  
the term ``2+0'' event sample becomes synonymous to the term ``2+0" bunch crossing 
sample.  

\begin{figure}[h]
\begin{center}
\setlength{\unitlength}{1mm}
\begin{picture}(160,70)
\put(0,0){\makebox(0,0)[lb]{
\epsfig{file=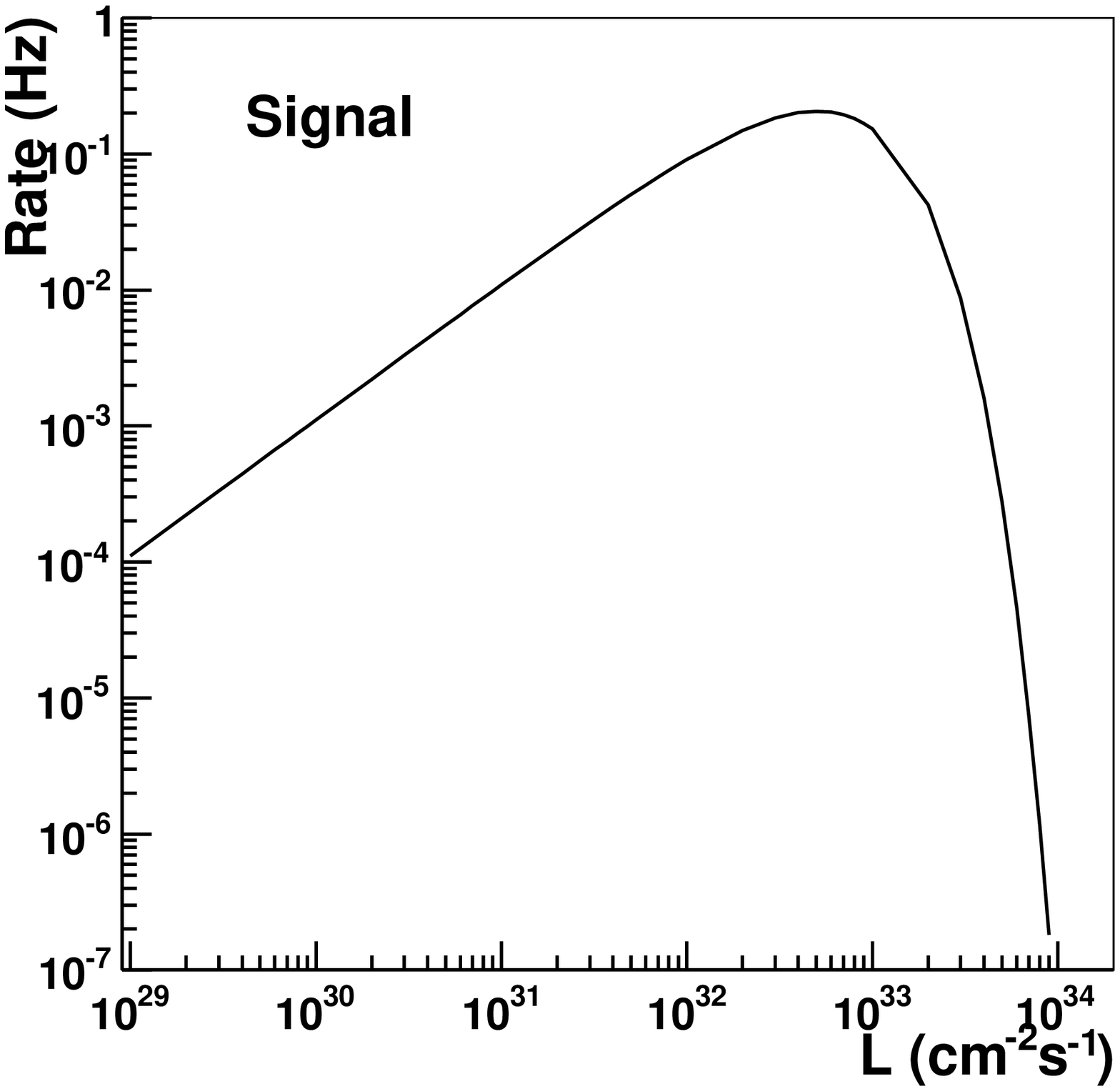,width=80mm,height=70mm}
}}

\put(75,0){\makebox(0,0)[lb]{
\epsfig{file=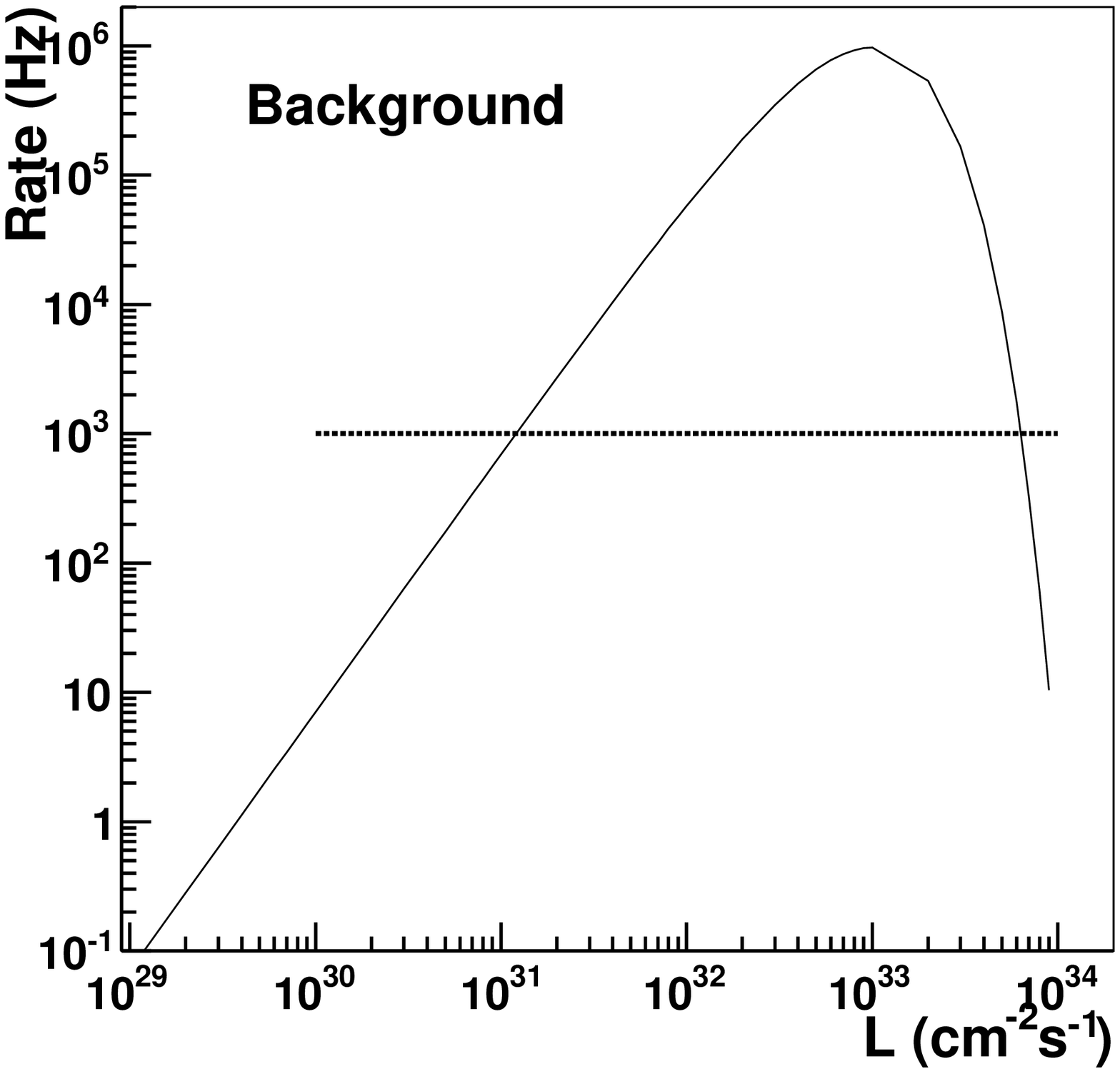,width=80mm,height=70mm}
}}
\put( 40,-5){\makebox(0,0)[cb]{\bf (a)}}
\put(115,-5){\makebox(0,0)[cb]{\bf (b)}}
\end{picture}
\end{center}
\caption{The rate ``2+0'' events  as a function of the machine luminosity:
(a) --  LPAIR signal events,
(b) --  PYTHIA background events.}
\label{plot2}
\end{figure}
 
In Figure \ref{plot2} the machine luminosity dependent rate of the ``2+0'' bunch 
crossings for a uniform bunch-by-bunch luminosity distribution and for the 25 ns 
bunch spacing is presented. Figure \ref{plot2}a shows the rate of  the LPAIR signal
events.  The rate of  the PYTHIA hadron  background
events is depicted in Fig. \ref{plot2}b.  The rate of the signal 
events reaches its maximum of 0.1 Hz at the luminosity
of $L=0.7\cdot 10^{33}$~cm$^{-2}$~s$^{-1}$ and decreases rapidly for the 
luminosity $L \geq 3\cdot 10^{33}$~cm$^{-2}$~s$^{-1}$ as the average number of 
the proton-proton collisions per bunch crossing increases and the 
probability of the silent bunch crossing decreases. The rate of the 
background events reaches the level of $700$ kHz at the luminosity
$L\approx~10^{33}$~cm$^{-2}$~s$^{-1}$ and decreases rapidly at higher
luminosity values for the same reasons. Figure \ref{plot2}a illustrates 
the instantaneous luminosity range over which a requisite statistical precision 
of the absolute luminosity measurement can be achieved, 
for a given data collection time interval.  As an example, for the data collected over one year the 
expected statistical precision is better than 1\% for the luminosity range 
$3\cdot 10^{30}\leq L\leq 3\cdot 10^{33}$~cm$^{-2}$~s$^{-1}$. Figure \ref{plot2}b 
defines the performance target for the luminosity detector LVL1 trigger. If one assumes 1~kHz input rate as the rate upper 
limit,  which will be allocated by the host  detector for  the luminosity detector LVL1 triggers,   an 
additional rejection factor of up to 700, must to be achieved. 

\subsection{Characteristics of the signal and background events}

%
\begin{figure}[h]
\begin{center}
\setlength{\unitlength}{1mm}
\begin{picture}(160,70)
\put(0,0){\makebox(0,0)[lb]{
\epsfig{file=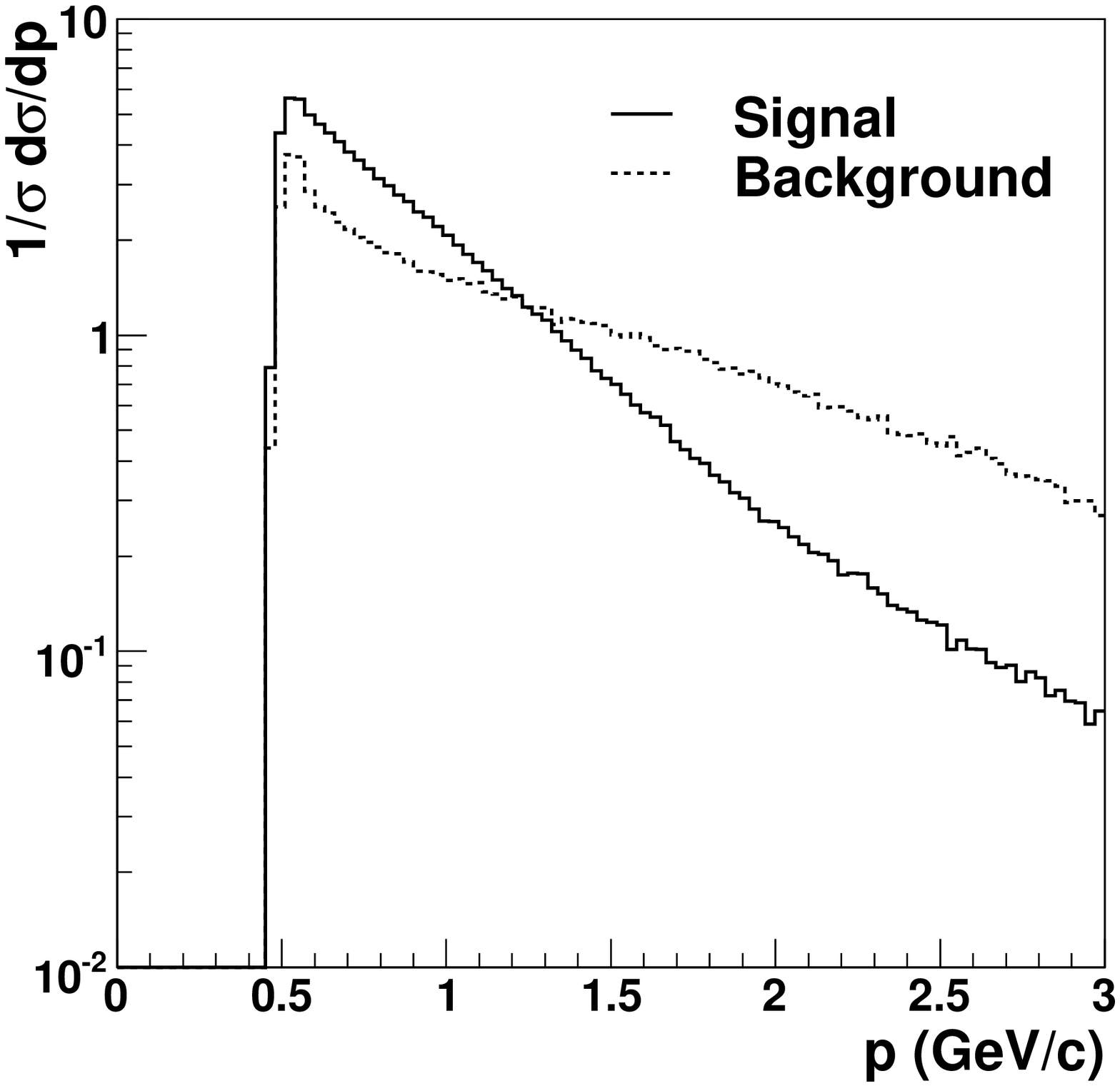, width=80mm,height=70mm}
}}
\put(75,0){\makebox(0,0)[lb]{
\epsfig{file=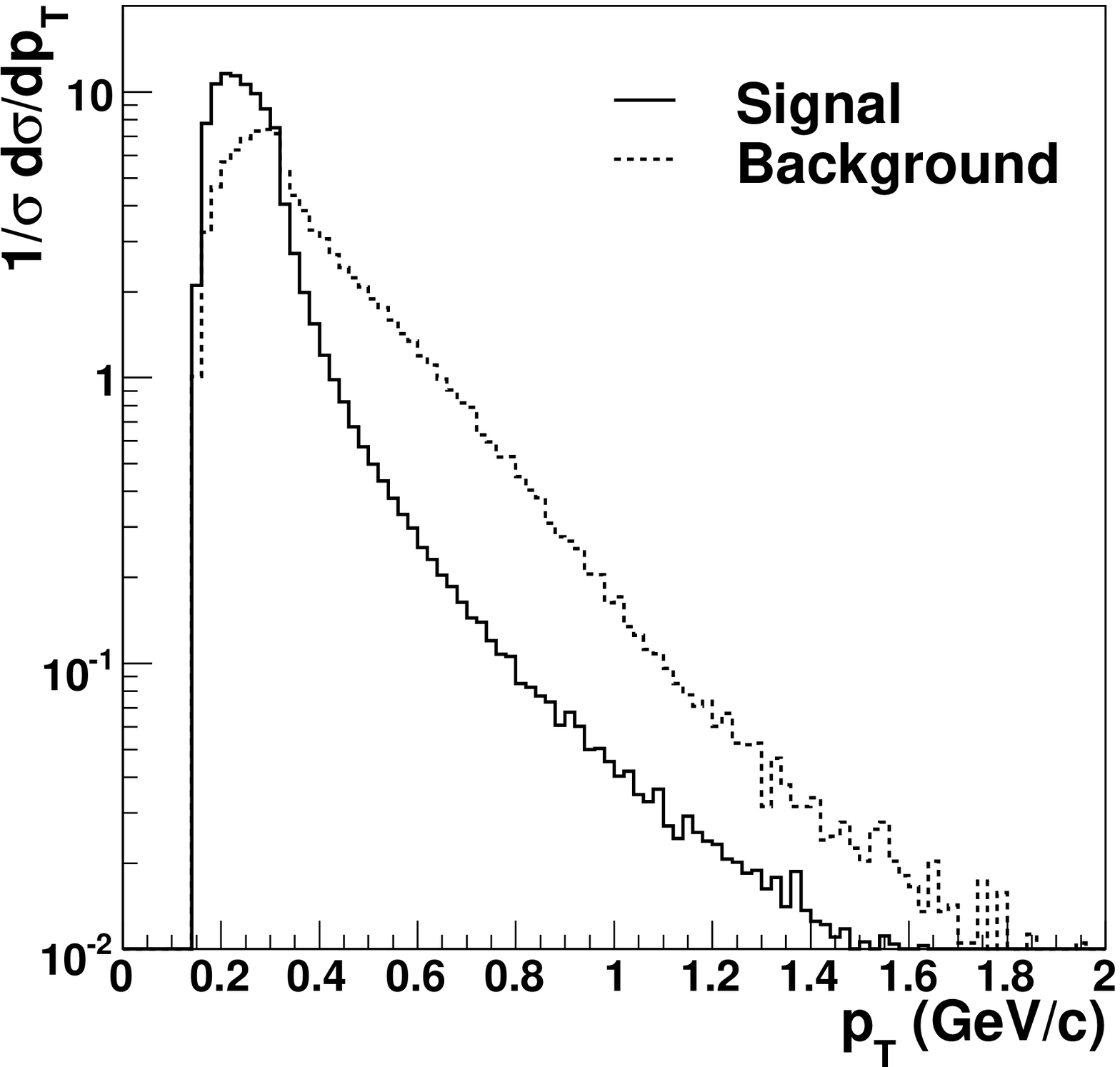, width=80mm,height=70mm}
}}
\put( 40,-5){\makebox(0,0)[cb]{\bf (a)}}
\put(115,-5){\makebox(0,0)[cb]{\bf (b)}}
\end{picture}
\caption{The distributions of:
(a)  -- the charged particle momentum,
(b)  -- the charged particle transverse momentum
for the  LPAIR signal (the solid line) and  the PYTHIA background (the dotted line) 
``2+0" events for the {\bf B2} field configuration.}
\label{plot3}
\end{center}
\end{figure}

In Figures \ref{plot3}a and \ref{plot3}b the distributions of the momentum, $p$, 
and of the transverse momentum, $p_T$, for the tagged particles in the ``2+0'' 
events are shown for the {\bf B2} configuration for the signal and background 
events, respectively. Two important observations are noted. Firstly, the 
momenta of the tagged particles are determined by  the Lorenz boost 
which, in turn, reflects the $z$ position of the luminosity detector fiducial volume. Secondly, both 
the $p$ and $p_T$ distributions are driven solely by the fiducial volume 
geometry and by the magnetic field strength. Consequently, the acceptance for the tagged 
particles will be nearly time-independent\footnote{Its 
time variation will be driven only by the evolution of the length of the 
proton bunches during the LHC luminosity run (the distributions shown in Figure 
\ref{plot3} correspond to the bunch size of 7.5 cm).}.
It is important to note that  the low transverse momentum 
particles will never reach radially the fiducial volume for the {\bf B2} configuration. 

\begin{figure}[h]
\begin{center}
\setlength{\unitlength}{1mm}
\begin{picture}(160,70)
\put(0,0){\makebox(0,0)[lb]{
\epsfig{file=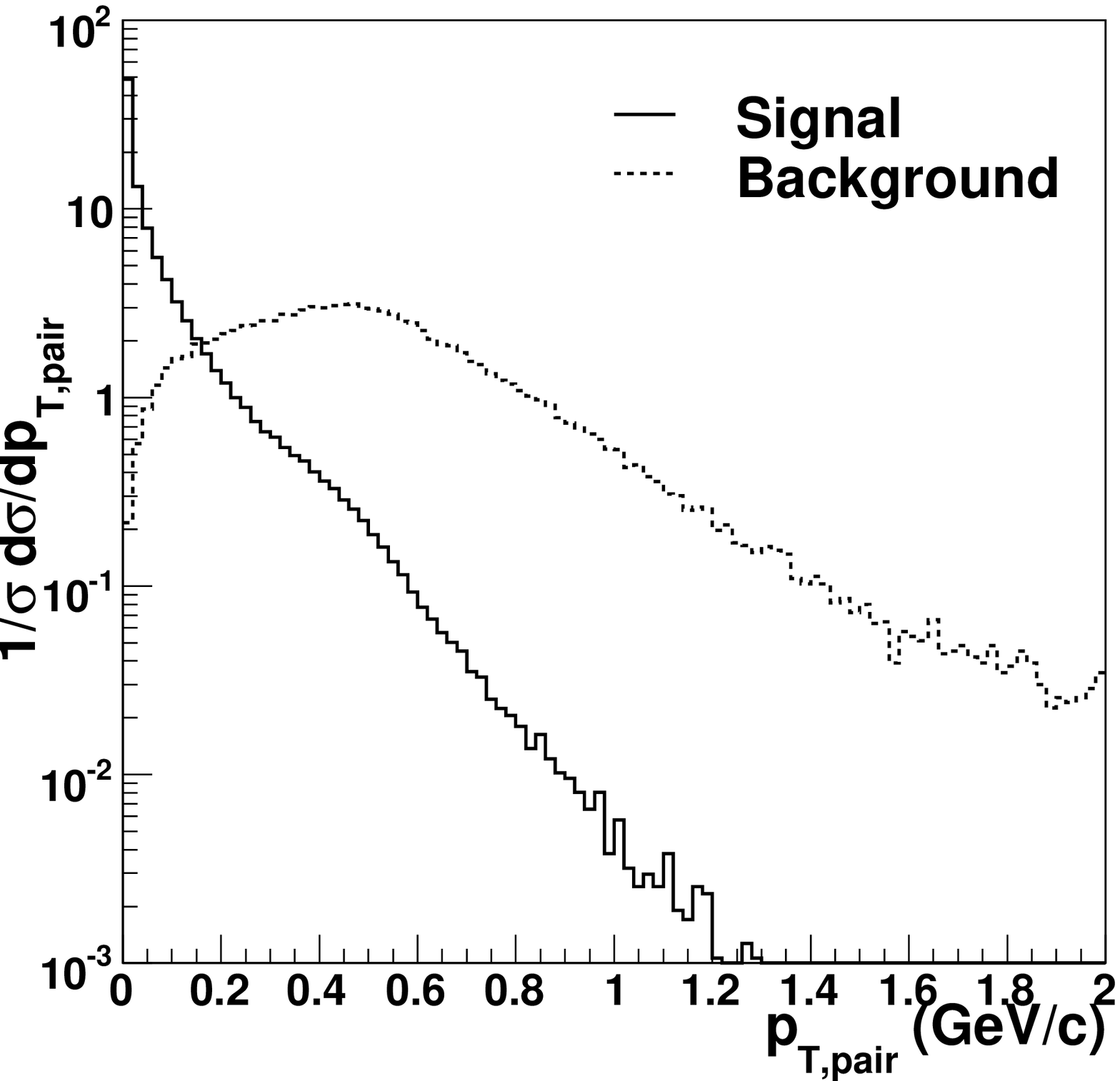, width=80mm,height=70mm}
}}
\put(75,0){\makebox(0,0)[lb]{
\epsfig{file=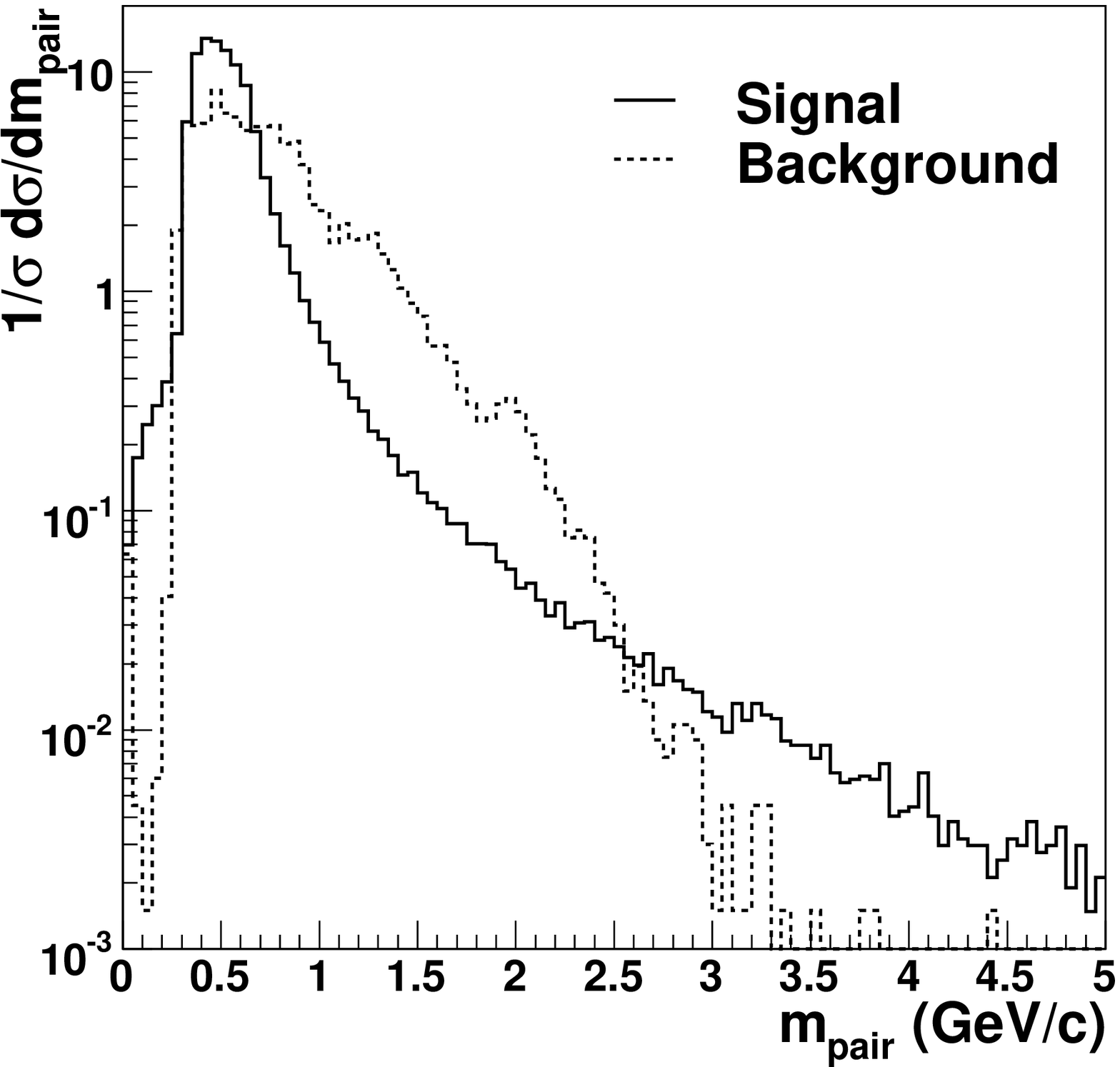, width=80mm,height=70mm}
}}
\put( 40,-5){\makebox(0,0)[cb]{\bf (a)}}
\put(115,-5){\makebox(0,0)[cb]{\bf (b)}}
\end{picture}
\caption{The distributions of: 
(a) -- the transverse momentum, 
(b) -- the invariant mass
for the unlike charge particle pairs for the LPAIR signal (the solid line) and the PYTHIA background
(the dotted line) ``2+0"  events for the {\bf B2} field configuration. }
\label{plot4}
\end{center}
\end{figure}

In Figures \ref{plot4}a and \ref{plot4}b the distributions of the transverse 
momentum, $p_{T,pair} = |\vec{p}_{T,1}+\vec{p}_{T,2}|$, of the two unlike charge 
tagged particles, and their invariant mass, $m_{pair}$, for the ``2+0'' signal 
and  background events are presented. The 
behaviour of $p_{T,pair}$ distribution in the limit of decreasing transverse momentum 
is different for the signal and the background samples. For the latter it 
decreases, as expected, to zero with decreasing $p_{T,pair}$ value while for the 
signal sample it peaks at small values of $p_{T,pair}$. This behaviour reflects 
the dominance of the point-like, electromagnetic coupling of protons to the 
virtual photons for the selected event sample (c.f. 
\cite{first} for a more detailed discussion of these aspects). The $p_{T,pair}$ variable could thus be used to 
select efficiently the signal events. Unfortunately, the requisite precision 
of the $p_{T,pair}$ reconstruction cannot be achieved within the 
LVL1 trigger latency and its  use has to be 
postponed to the subsequent HLT stages of the data selection procedure.

The distribution of the invariant mass of the pair of tagged particles reflects 
the corresponding distribution of the pair transverse momentum. The back-to-back 
pairs have, in general, larger masses for the same momentum 
spectrum of each of the tagged particles. Note the presence of the hadron  
resonances (for example $f2(1270)$) and of the  barion-antibarion pairs (in the vicinity 
of 2 GeV) in the invariant mass distribution for  the background pairs. Note as well  
the contribution of the Dalitz decays of neutral pions visible in the region of 
the smallest invariant masses. The particle type decomposition of the background sources in 
various mass regions will be crucial for the  data driven background subtraction scheme. 

The measure of the back-to-back topology of a pair of charged particles is the 
acoplanarity \cite{first}, $\delta\phi$, defined as:
$$\delta\phi=\pi-min(2\pi-|\phi_{1}-\phi_{2}|,|\phi_{1}-\phi_{2}|)$$ 
where $\phi_{1}$ and $\phi_{2}$ are the azimuthal angles of the
particles at the interaction vertex\footnote{To be precise, 
the variable  defined above describes  the acollinearity of the 
 transverse momenta vectors of the particles belonging to a pair.
Its name, even if misleading,  is retained in our series of papers to follow the convention of the corresponding literature.}. 
The reduced acoplanarity, $\delta\phi_r$, with
values between 0 and 1 is defined as $\delta\phi_r = \delta\phi/\pi$.

\begin{figure}[h]
\begin{center}
\setlength{\unitlength}{1mm}
\begin{picture}(160,70)
\put(0,0){\makebox(0,0)[lb]{
\epsfig{file=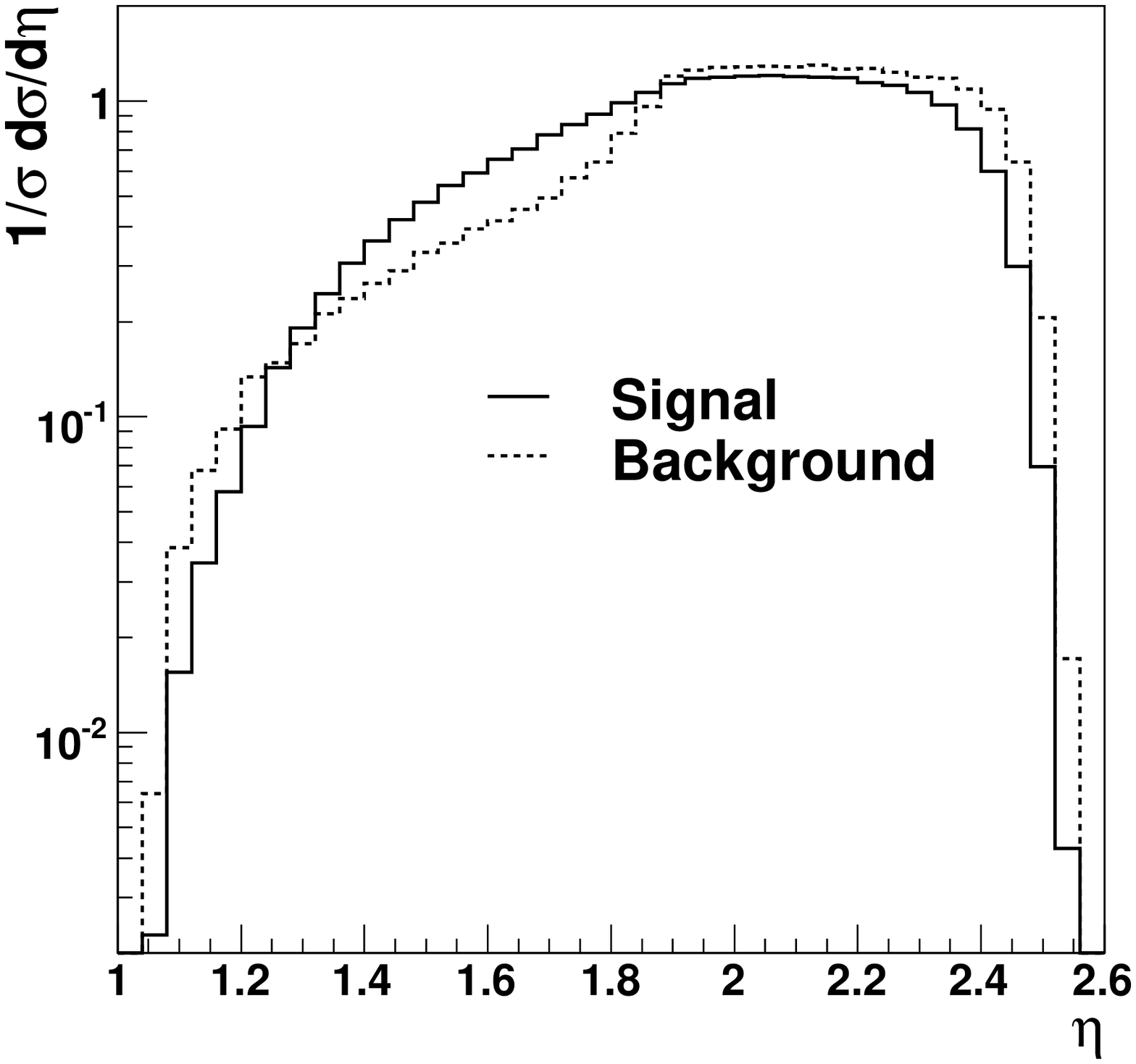, width=80mm,height=70mm}
}}
\put(75,0){\makebox(0,0)[lb]{
\epsfig{file=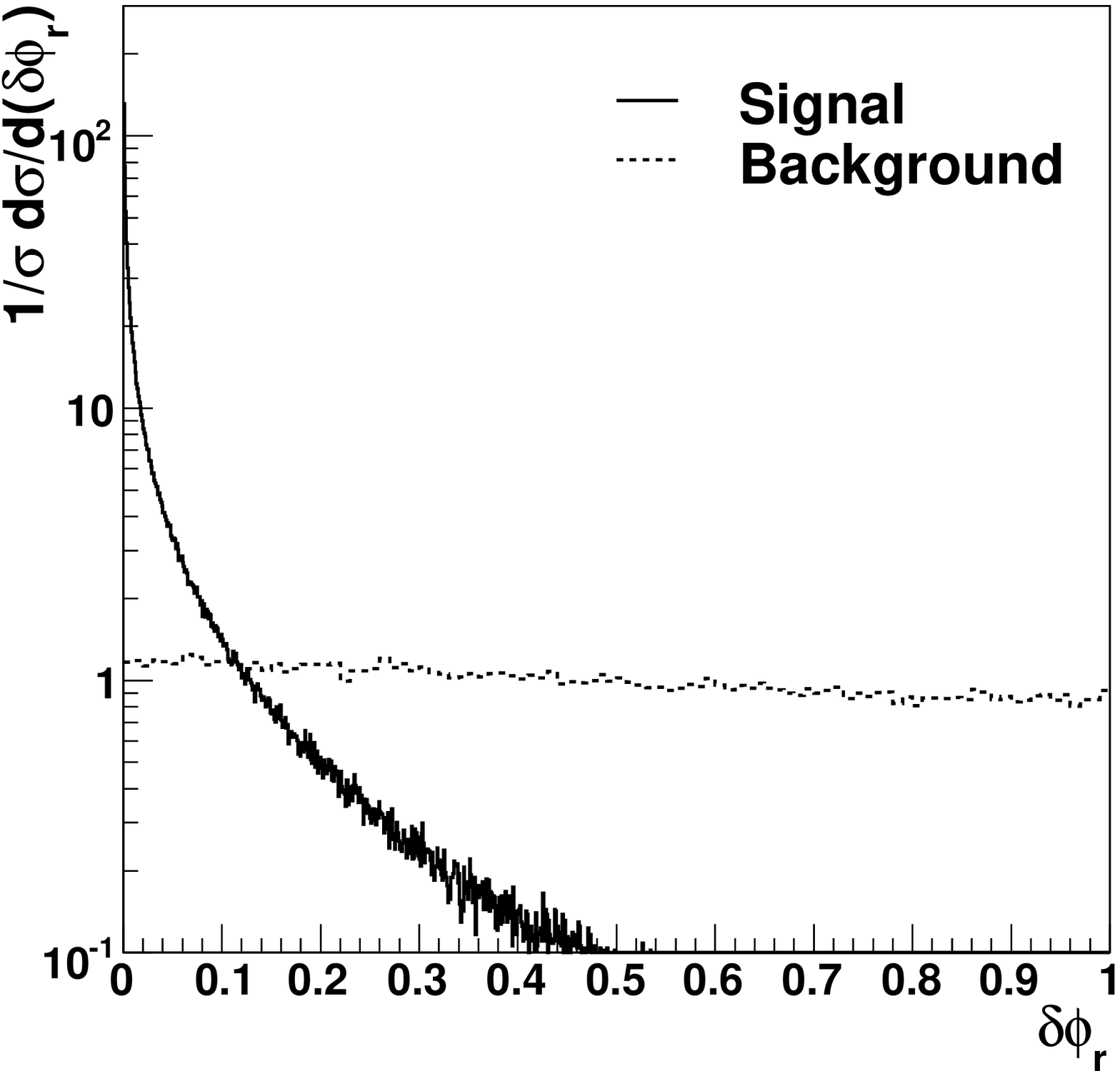, width=80mm,height=70mm}
}}
\put( 40,-5){\makebox(0,0)[cb]{\bf (a)}}
\put(115,-5){\makebox(0,0)[cb]{\bf (b)}}
\end{picture}
\end{center}
\caption{The distributions of:
(a) -- the charged particles pseudorapidity, $\eta$,
(b) -- the unlike charge particle pair reduced acoplanarity, $\delta\phi_r$,
for the LPAIR signal (the solid line) and the PYTHIA background (the dotted line) 
``2+0'' events for the {\bf B2} field configuration. }
\label{plot5}
\end{figure}

In Figs. \ref{plot5}a and \ref{plot5}b the distributions of the particle
pseudorapidity, $\eta > 0$, and the reduced acoplanarity, $\delta\phi_r$, of the 
unlike charge tagged particle pairs in ``2+0'' events  are drawn, respectively. A broad $\eta$ distribution, similar for 
the signal and the background events, illustrates the effect of the magnetic
field which enlarges the acceptance region for particles of the relatively small pseudorapidity.
The reduced acoplanarity distribution peaks strongly at $\delta\phi_r\sim 0$ for the
signal events. It is approximately flat for the background events.  This
difference will be of principal importance for the discrimination of the 
background against the signal by the luminosity detector LVL1 trigger logic. Note, that 
the flat shape of the reduced acoplanarity distribution for the background events, 
being  insensitive to the details of the modelling of the hadron interactions
(it reflects merely the longitudinal phase-space),  is perfectly suited  for a 
precise, data-based estimation of the background contribution to the signal 
peak.

An efficient and fast selection of the ``2+0'' tagged particle configuration and a precise reconstruction of the particle pair acoplanarity will drive 
the luminosity detector the LVL1 trigger performance requirements 
discussed in the next section. 

\section{The detector performance requirements}
\label{sec:detperf}

\subsection{Timing resolution}

The principal requirement for the timing resolution of the luminosity detector 
signals is their unambiguous association to the appropriate bunch crossings by the 
LVL1 trigger logic. Only if this requirement is met the silent bunch crossings 
can be unambiguously attributed (hence synonymous) to ``no-luminosity-detector-signal'' 
bunch crossings\footnote{Note, that in principle one could use the notion of the 
``silent-group-of-bunch crossings'' if the detector signals cannot be assigned 
unambiguously to a single bunch crossing. In such a case the applicability domain of the 
discussed method will be restricted only to the low luminosity machine 
operation.}.

In our view, a  robust  bunch crossing association of the 
detector signals for the {\bf low momentum} tagged particles must be  
based on the luminosity detector  {\bf LVL1 track segment candidates}. 
The LVL1 timing resolution of the track 
segments depends both upon the dispersion of the particle flight time before  
reaching  the fiducial volume of the luminosity detector and upon the intrinsic timing  
resolution of the luminosity detector signals. A 
robust LVL1 trigger must not only unambiguously assign  the track segment signal to  a 
correct bunch crossing but,  in addition,  it should efficiently reject spurious
beam halo and random detector noise track candidates. 
\begin{figure}
\begin{center}
\setlength{\unitlength}{1mm}
\begin{picture}(160,150)
\put(0,75){\makebox(0,0)[lb]{
\epsfig{file=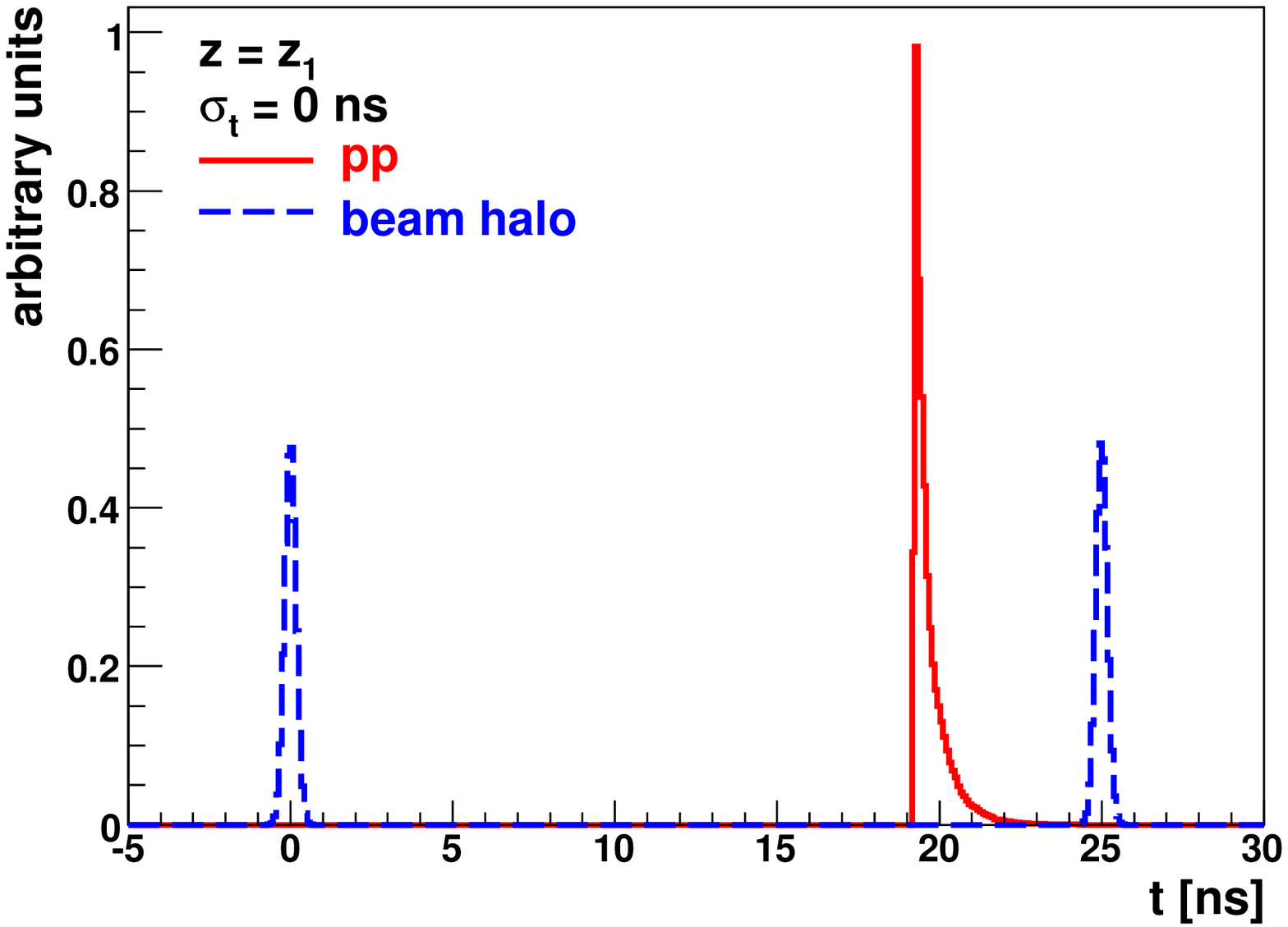, width=80mm,height=70mm}
}}
\put(0, 0){\makebox(0,0)[lb]{
\epsfig{file=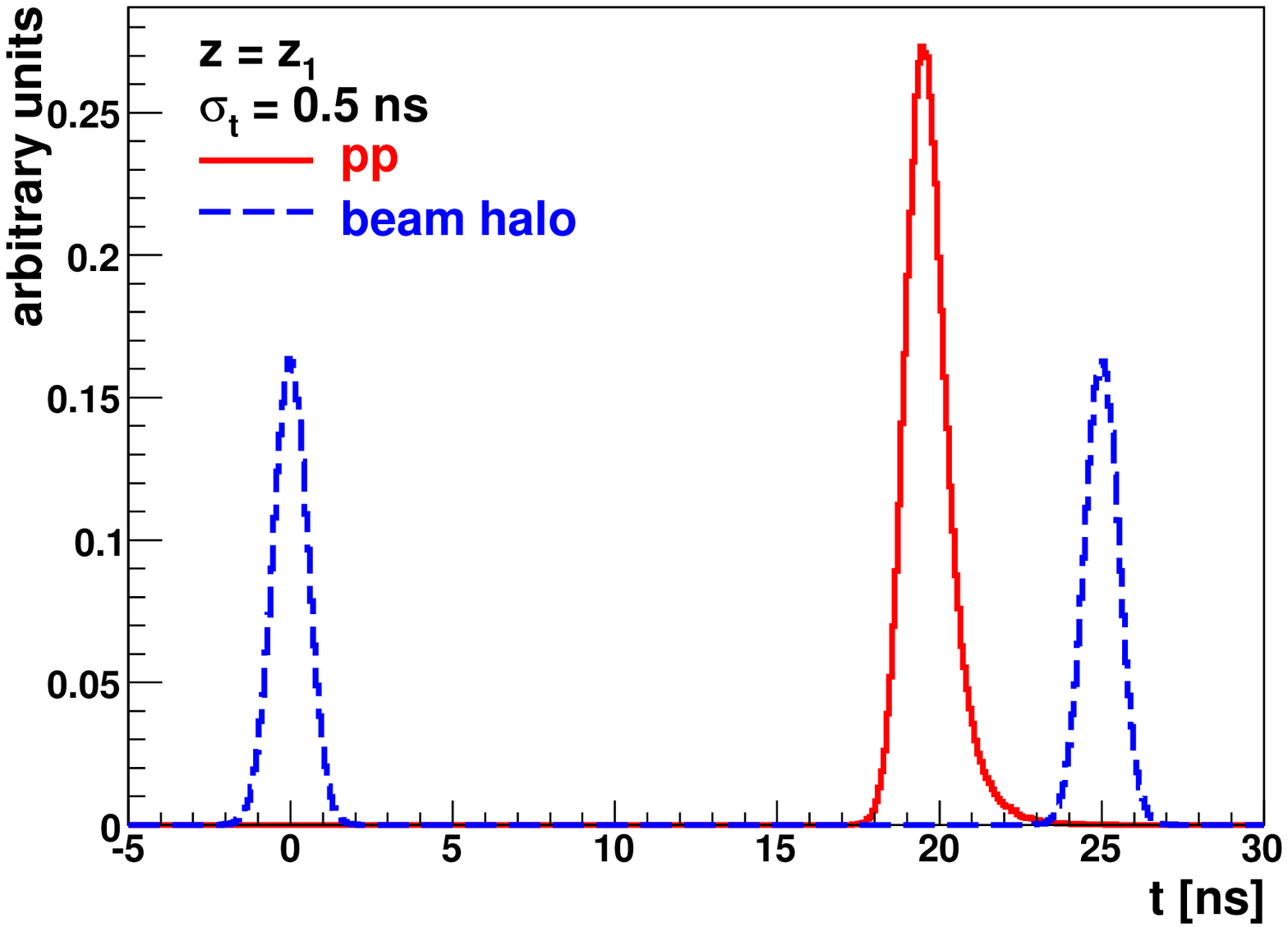, width=80mm,height=70mm}
}}
\put(75,75){\makebox(0,0)[lb]{
\epsfig{file=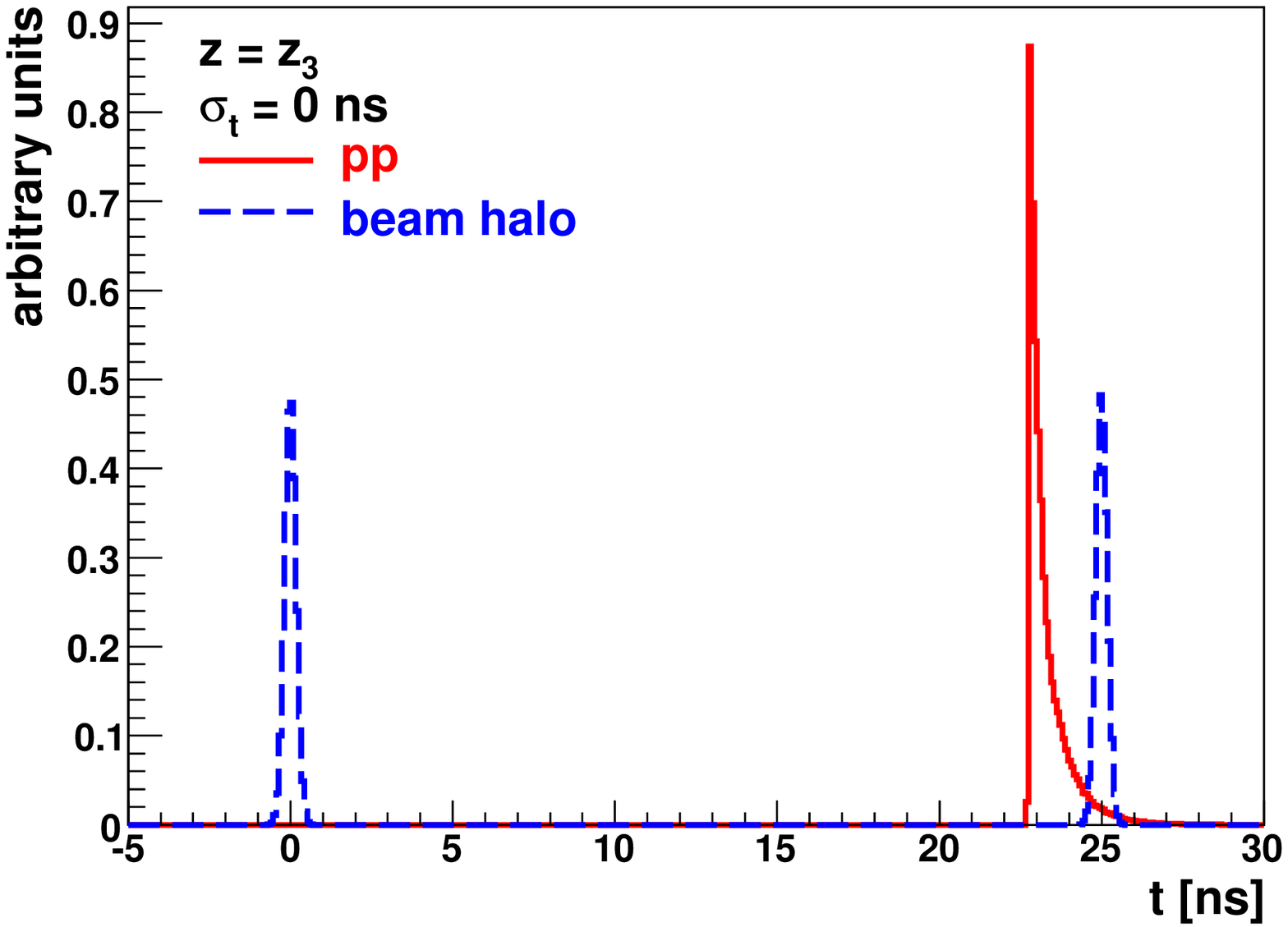, width=80mm,height=70mm}
}}
\put(75, 0){\makebox(0,0)[lb]{
\epsfig{file=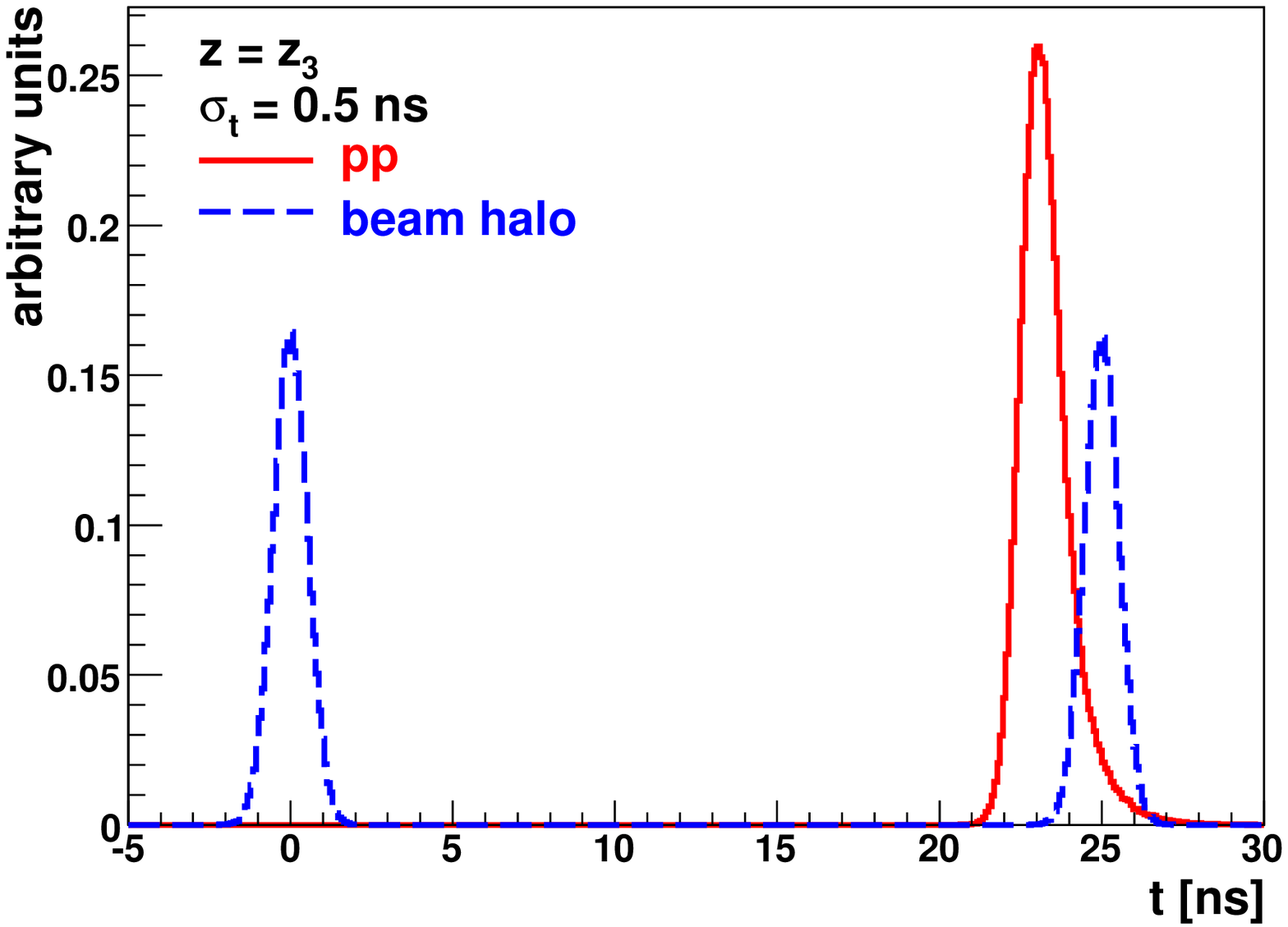, width=80mm,height=70mm}
}}
\put( 40,75){\makebox(0,0)[cb]{\bf (a)}}
\put(115,75){\makebox(0,0)[cb]{\bf (b)}}
\put( 40,-5){\makebox(0,0)[cb]{\bf (c)}}
\put(115,-5){\makebox(0,0)[cb]{\bf (d)}}

\end{picture}
\end{center}
\caption{The distributions of the particles' arrival time at:
(a) -- $ z = z_1$, (b) -- $z = z_2$ planes. The beam-beam collision vertex
originated particles are represented by the full line while the beam
halo particles are represented by the dashed line. The relative 
normalisation of the above two distributions will depend upon 
the beam conditions and is, at present, arbitrary.  The effect of the
0.5 ns Gaussian smearing of arrival-time measurement is shown in plots (c) and (d).}
\label{plot6}
\end{figure}
In Figures \ref{plot6}a and \ref{plot6}b we show the distributions of the 
arrival times of the particles at the $z_1$ and $z_3$ planes, respectively. The 
distribution for the tagged particles originating from the bunch-bunch collision 
vertex is marked with the solid line. The Gaussian distribution of the $pp$ collision  
time within a bunch crossing of 180 ps was assumed. The distributions of the 
arrival time at the $z_1$ and $z_3$ 
detector planes for  the beam halo particles  are plotted with the broken line. 
It was assumed that the halo 
particles move in phase (synchronously) with proton bunches. They enter first the $z_3$ 
and then the $z_1$ plane. The tagged particles produced in the bunch-bunch 
collisions cross the first (third) detector plane delayed by $\sim 20$($\sim 
23$) ns with respect to the halo particles associated with the interacting
bunch. They arrive to these planes $\sim 5$($\sim 2$) ns before the
halo particles associated with the subsequent bunch (for 25~ns bunch crossing 
spacing). The arrival time distribution for the tagged particles has a 
significant tail of the ``late particles''  -- specific for the ${\bf B2}$ magnetic 
field configuration.  These ``late particles'' are characterised by a long helix trajectory. 

This study was repeated assuming that the intrinsic detector time resolution 
of the hits  measured  in the $z_1$ and $z_3$ planes is $0.5$ ns. Results of the 
calculations are presented in Figs. \ref{plot6}c and \ref{plot6}d. These plots
demonstrate that the beam halo particles can be efficiently discriminated 
against the tagged particles originating from the bunch-bunch collision using 
the measurement of the particle arrival time at  the $z_1$ plane, provided that the intrinsic hit 
timing resolution is not worse than 0.5~ns. For the $z_3$ plane position the halo 
particle hits and the $pp$ collision associated hits cannot be resolved  if the machine
operates with the 25~ns bunch spacing mode\footnote{The precise 0.5~ns resolution timing 
of  the detector signals  
available for the LVL1 trigger could be of use not only for the luminosity 
measurement but also for an efficient veto against any high energy halo muons
interacting in the detector material and mimicking the missing transverse energy 
LVL1 signatures. Another important, potential gain from an improvement of the 
timing resolution is the capacity of a precise discrimination of the 
proton/kaon/pion track segments. It could be of importance for tagging 
heavy flavour particles.}.

\subsection{Spatial resolution}

The discussion presented below is restricted to the requirements for measurement 
resolution of the azimuthal angle, $\phi$, of the hits left by the tagged 
particles traversing the luminosity detector planes. The angular positions of the 
hits in the $z_1, z_2$ and $z_3$ detector planes will be denoted by 
$\phi_1, \phi_2$ and $\phi_3$, respectively. A measurement of the radial 
positions of the hits would certainly be useful in rejecting the 
spurious track candidates but it is not indispensable for 
the proposed method.  Thus, the radial segmentation of the fiducial volume will 
not be discussed here.

The presence of the hits in the three detector planes is a minimal necessary 
condition to detect the interaction-vertex-unconstrained track segments  in the 
case of both the {\bf B2} and the {\bf B0} field configurations.  The required 
$\phi$ hit resolution in each of the $z$ planes will be determined  by the requirement of 
the reduction of the LVL1 accepted rate of the ``2+0'' tagged particle pairs to 
a level of $\sim$1~kHz,  while retaining a large fraction of the lepton pairs 
produced in the kinematic region of small acoplanarities. 

To quantify the detector resolution requirements  we introduce the {\bf coplanar pair 
selection efficiency} estimator, $\epsilon(\sigma_\phi)$. Its dependence  upon the measurement 
resolution of the azimuthal hit position, $\sigma_\phi$, is given by the following formula:  
\begin{equation}
\epsilon(\sigma_\phi) = \frac
{\int_{0}^{0.01}f(\delta\phi_r^{rec},\sigma_\phi)d\delta\phi_r^{rec}}
{\int_{0}^{0.01}f(\delta\phi_r,0)d\delta\phi_r}
\label{eq-effi}
\end{equation}
where:  $\delta\phi_r^{rec}$ is the reconstructed (reduced) pair acoplanarity, and  
$f(\delta\phi_r,0), f(\delta\phi_r^{rec},\sigma_\phi)$ are, respectively,  the true 
and the reconstructed reduced acoplanarity 
distributions\footnote{The restriction of the reduced acoplanarity range is 
very handy for the detector requirements quantification. 
The rate of the leptons pairs 
in this range can be controlled theoretically with a precision better 
than 1\% \cite{first}. The anticipated reduction of the  background rate is of 
the order of a factor 100 due to approximate flatness of the acoplanarity 
distribution for the background events (c.f. Fig. \ref{plot5}b).}.

This estimator will be used to map  the ``compromise space"  between the 
requisite luminosity detector granularity (determining its costs) and the LVL1 trigger
capacity for selection of coplanar lepton pairs.  

\subsubsection{The $B0$ field configuration}

In the case of the {\bf B0} field configuration the average azimuthal position 
of the hits in the three detector planes, $<\phi> = 1/3\cdot(\phi_1+\phi_2+\phi_3)$, 
provides the best estimate  of the azimuthal angle of the tagged particle and 
hence, the reconstructed  acoplanarity, $\delta\phi_r^{rec}$,   of a pair.
 
\begin{figure}[t]
\begin{center}
\setlength{\unitlength}{1mm}
\begin{picture}(160,70)
\put(0,0){\makebox(0,0)[lb]{
\epsfig{file=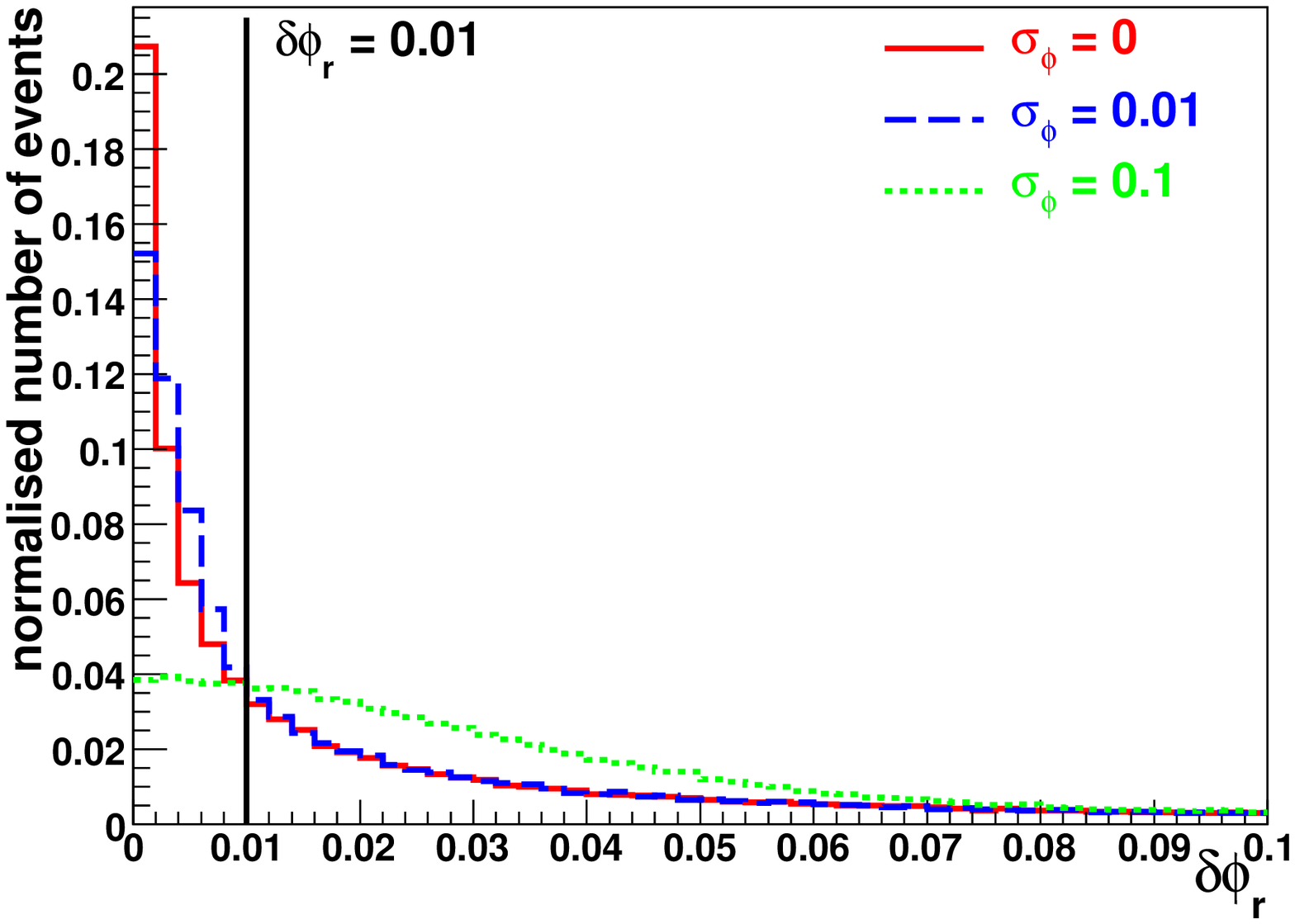, width=80mm,height=70mm}
}}
\put(75,0){\makebox(0,0)[lb]{
\epsfig{file=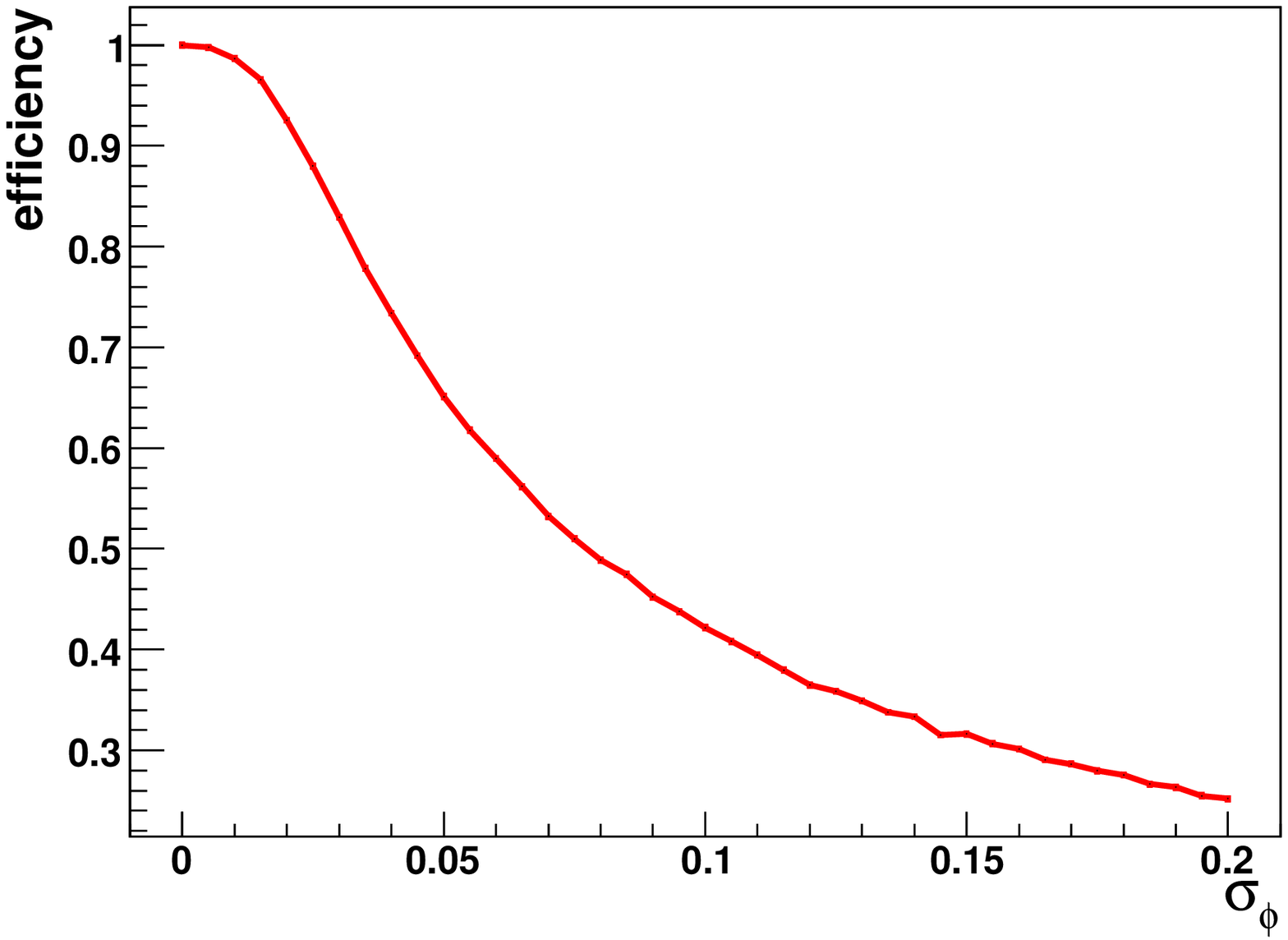, width=80mm,height=70mm}
}}
\put( 40,-5){\makebox(0,0)[cb]{\bf (a)}}
\put(115,-5){\makebox(0,0)[cb]{\bf (b)}}
\end{picture}
\end{center}
\caption{The resolution studies for the $B0$ field configuration (LPAIR signal events):  
(a) -- the reduced acoplanarity distribution 
for the three values of the detector azimuthal resolution,
$\sigma_\phi$: 0.0 (the solid line histogram), 0.01 (the dashed line  histogram) and 0.1 (the doted line).  
(b) -- the efficiency, $\epsilon(\sigma_\phi)$ (see text), as a function of the
detector azimuthal resolution, $\sigma_{\phi}$.
}
\label{plot7}
\end{figure}

In Fig. \ref{plot7}a the $\delta\phi_r^{rec}$ distribution of the lepton pairs for the three 
values of the detector Gaussian azimuthal resolution, $\sigma_\phi$, is shown.  
The distribution  changes with deteriorating resolution of the 
detector.  For $\sigma_\phi = 0.1$ the characteristic peak at $\delta\phi_r \simeq 0$ 
disappears due to the migration of events towards the higher acoplanarity 
values.  In Fig. \ref{plot7}b the efficiency, $\epsilon(\sigma_\phi)$, as a 
function of the azimuthal detector resolution, $\sigma_\phi$ is presented. The 
efficiency was calculated assuming an  absence of any detector effects other 
than the detector resolution. The efficiency drops from value of 1 for a perfect 
detector to about 0.25 if the particle hit  azimuthal resolution is 0.2 radians. For 
the ${\bf B0}$ field configuration a detector with the azimuthal hit resolution of 
about 0.03 radians would be sufficient to select the coplanar lepton pair 
production events with the efficiency better than 0.9 while reducing the
hadronic background rate by a factor of about 100.

\subsubsection{The $B2$ field configuration}

In the presence of the solenoidal magnetic field the initial acoplanarity of the 
small-transverse-momentum lepton pairs will no longer be reflected in the back-to-back topology of the 
luminosity detector track segments. Their topology depends on 
the pair invariant mass and rapidity,  and on  the emission angles of a positively (negatively) charged 
lepton in the pair rest frame. A multidimensional unfolding of the complete set of lepton pair kinematic 
variables is fairly complicated and, thus, of little use for efficient LVL1 event selection. The 
trick proposed here is to directly project  the topology of the particles' hits onto the initial (interaction vertex)
pair acoplanarity ignoring the complete reconstruction  of the pair kinematics.

In absence of the dead material on the particle path from the interaction vertex to the luminosity detector 
fiducial volume the charged particles move with a constant azimuthal velocity in 
the plane perpendicular to the B field, the $(x,y)$ plane, and also with a 
constant velocity along the B field (the $z$ axis direction). Using cylindrical 
coordinates the particle azimuthal position, $\phi$, evolves with time as:
\begin{equation}
\phi(t) =  \phi(z_{vert}) + \omega (B, p_T) \cdot (t - t_{vert}),
\label{eq-phi}
\end{equation}
and its position along the z axis as:
\begin{equation}
z(t) =  z_{vert} + v_z (p, p_T) \cdot (t - t_{vert})
\label{eq-z}
\end{equation}
where $z_{vert}$ and $t_{vert}$ are the $z$ position of the pair production
vertex and the $pp$ collision time, respectively, $\omega(B, p_T)$ denotes the tagged 
particle angular velocity and $v_z(p, p_T)$ is the $z$ component of the particle 
velocity.

In absence of the $z_{vert}$ and $t_{vert}$ smearing effects (for ``pancake"-like bunches)
the pair acoplanarity could, in principle,  be determined from the above equations
using the measured azimuthal positions of the hits left by the tagged particles.  In reality,
the $\delta\phi_r$ reconstruction  must be done simultaneously with the unfolding of 
the interaction vertex,  $z_{vert}$,  position. 

\begin{figure}[ht]
\begin{center}
\setlength{\unitlength}{1mm}
\begin{picture}(160,70)
\put(45,0){\makebox(0,0)[lb]{
\epsfig{file=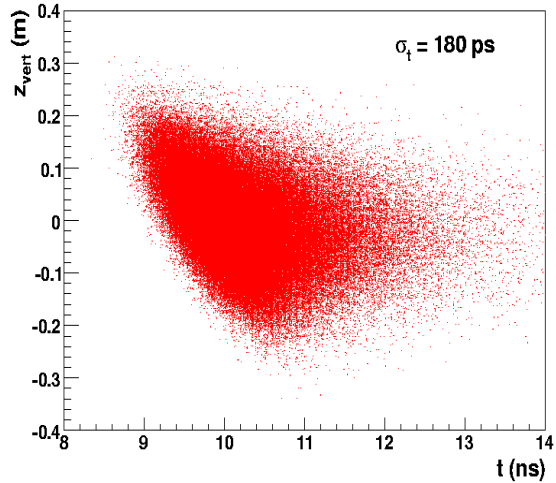,width=80mm,height=70mm}
}}
\end{picture}
\end{center}
\caption{The correlation between the interaction vertex $z_{vert}$ coordinate 
and the arrival time, $t_1$, of the tagged particles (LPAIR signal events)
to the detector $z_1$  plane. The Gaussian bunch density profile with
$\sigma = 7.5$ cm has been assumed.}
\label{plot8}
\end{figure}
In Fig. \ref{plot8} the correlation of the interaction vertex longitudinal coordinate, $z_{vert}$, and the time, $t_1$, required for the tagged particle to arrive at the $z_1$ detector plane is shown.  The Gaussian shape of the longitudinal bunch density profile has been assumed with the dispersion of $7.5$~cm.  At $t = 0$ the interacting bunches fully overlap at $z = 0$ position. Two effects influence the  $t_1$ time for a given vertex position. The first
one is the jitter of the collision time determined by the bunch size. It 
is of the order of 180 ps. The second, reflects the distribution of  the particles' time-of-flight  
from the production vertex to the $z_1$ plane. The time-of-flight  depends on the particle type, 
momentum,  and its production angle. For the momentum range discussed in this 
paper the tagged particles move with speed of light along a helix trajectory. Thus  
the time delay of a hit depend only upon the helix length. Figure \ref{plot8} 
shows that the correlation between the particle arrival time and the vertex 
position z coordinate is weak and that the measurement of $t_1$ can hardly constrain the vertex position.

A minimal detector requirement to reduce the vertex position uncertainty on the 
event-by-event basis and, as a consequence, to improve the precision of the initial pair acoplanarity 
reconstruction, is to measure not only the arrival times of particles 
at the $z_1$ plane but also the time-of-flight  
between the $z_1$ and $z_3$ planes.

The equations of motion (\ref{eq-phi}) and (\ref{eq-z}) written for the two 
tagged particles have 8 unknown parameters: the interaction vertex $z$-position, the collision time, the angular and axial velocities and the initial azimuthal angles of each of the two particles. These parameters can be, in principle, unfolded on the pair-by-pair basis from the measured $\phi$ positions and the relative time of the hits in any two detector planes. However, in reality, the equations of motion along the $z$ axis are quasi-degenerated for the relativistic particles and unrealistic precision of the time measurement would be required to
solve this system of equations. Therefore, two approximate reconstruction methods of the $z_{vert}$ vertex position and the reduced  (vertex) acoplanarity of a pair are proposed and evaluated in the following.

In the first method the $pp$ collision time jitter is neglected and it is assumed that all collisions take place at $t = 0$. 
In such a case the $z$-position of the vertex can be estimated, for the $i^{th}$ particle using the time of the hits, $t_1$ and $t_3$ in the $z_1$ and the $z_3$ plane, respectively, as:
\begin{equation}
z_{vert}^i =z_1 - \frac{z_3-z_1}{t_3 ^i - t_1^i} \cdot t_1^i.
\label{eq-zest_1}
\end{equation}
The interaction vertex position is then taken as the average of the positions calculated for each of the tagged particles:
\begin{equation} 
z_{est1} = \frac{z_{vert}^1 +z_{vert}^2}{2}.
\end{equation}
The second method is based on the following measured quantities, for each of the tagged particles denoted by a superscript $i$:
\begin{itemize}
\item the arrival time to the first detector plane, $t_1^i$,
\item the angular distance, $\Delta\phi_{31}^i = \phi_3^i - \phi_1^i$, 
of the particle hits in the $z_3$ and $z_1$ detector planes,
\item the time-of-flight, $t_3^i-t_1^i$ of  a particle moving between the $z_1$ and  
$z_3$ planes.
\end{itemize}
This method simultaneously unfolds the collision time and the position of the 
vertex under the assumption that the two  particles  are coplanar (back-to-back in the 
transverse plane). Such an additional assumption removes  the 
degeneracy of the system of eight linear equations\footnote{Note, that the 
system of eight linear equation is invariant with respect to $\phi$ rotation of 
the reference frame. Therefore,  the above six measured quantities and one external 
constraint is sufficient to reconstruct fully the particle pair kinematics.}. This method 
provides, thus, a precise estimation of the vertex position for the coplanar pairs. 
It biases, however,  the reconstructed vertex position for the acoplanar pairs. The 
resulting deterioration of the resolution is regularised by taking 
the $z = 0$ position for all the pairs for which the estimated collision time is 
outside the $\pm$360~ps wide  interaction time window. In the following, the vertex 
position reconstructed with this method will be denoted as $z_{est2}$.

The above two reconstruction methods are  compared,  in the 
following,  with a direct method in which  no attempt
to reconstruct the position of the vertex is made and all pairs are
assumed originate from the vertex fixed at $z_{est0} = 0$. 
In the studies of the relative precision of the above three vertex reconstruction methods it was 
assumed that the hit-timing in the $z_1$ plane is measured with a Gaussian resolution  of 100~ps while 
the relative time difference between the hits in the $z_1$ and $z_3$ planes is
measured with 20~ps accuracy. The calculations were carried out for the 7.5 cm 
long proton bunches with a Gaussian distribution of their intensities.

\begin{figure}
\begin{center}
\setlength{\unitlength}{1mm}
\begin{picture}(160,150)
\put(0,75){\makebox(0,0)[lb]{
\epsfig{file=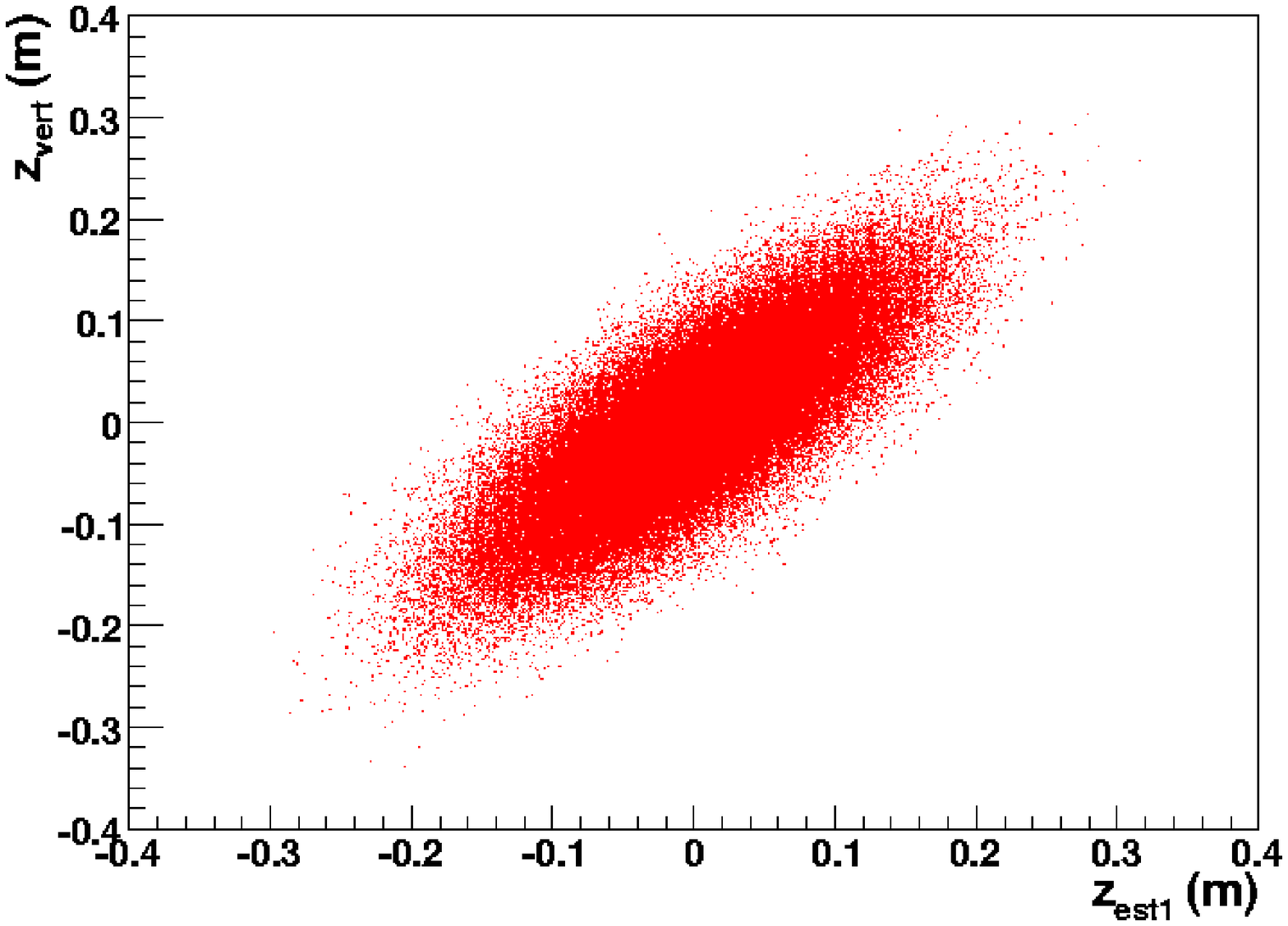, width=80mm,height=70mm}
}}
\put(0, 0){\makebox(0,0)[lb]{
\epsfig{file=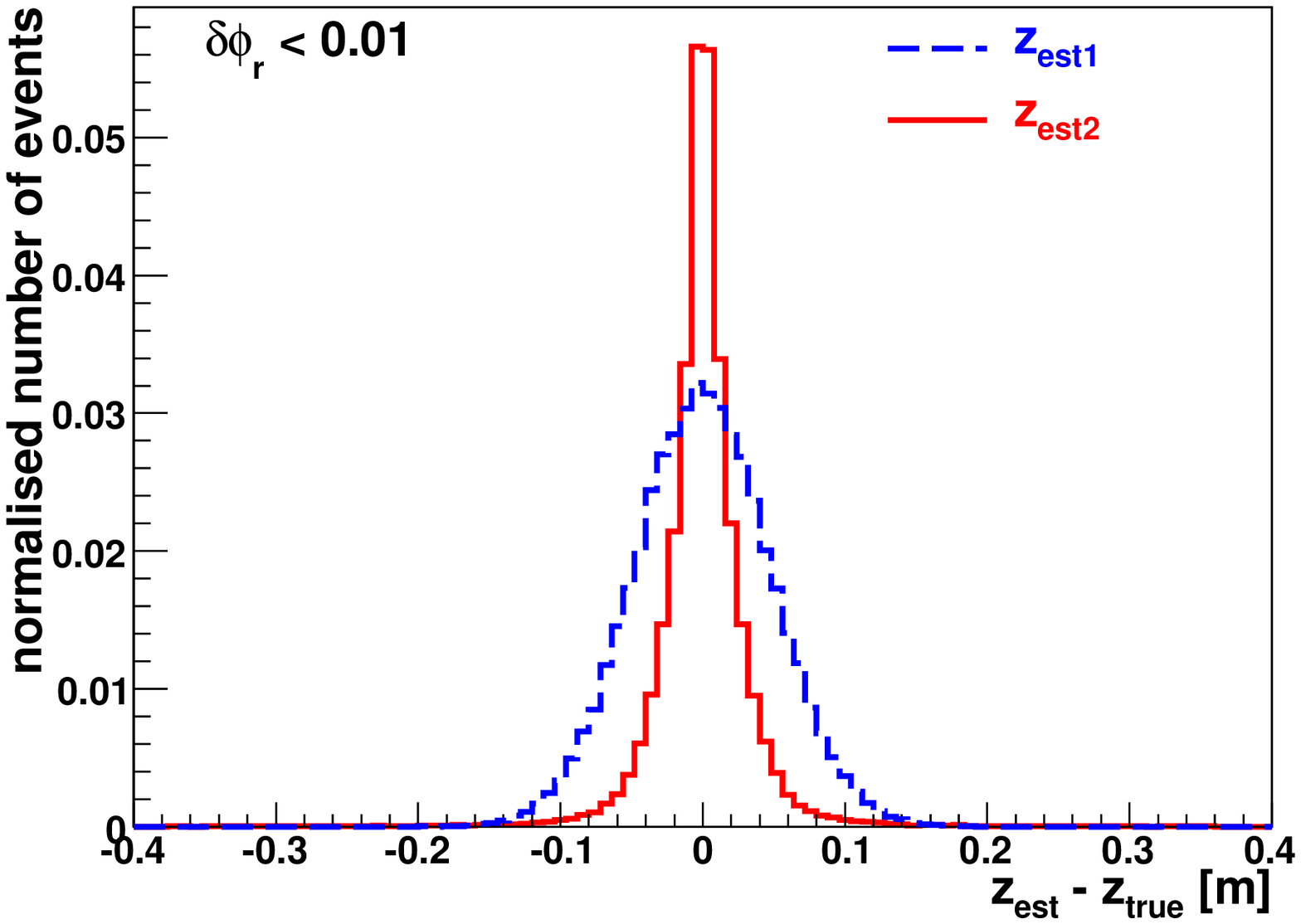, width=80mm,height=70mm}
}}
\put(75,75){\makebox(0,0)[lb]{
\epsfig{file=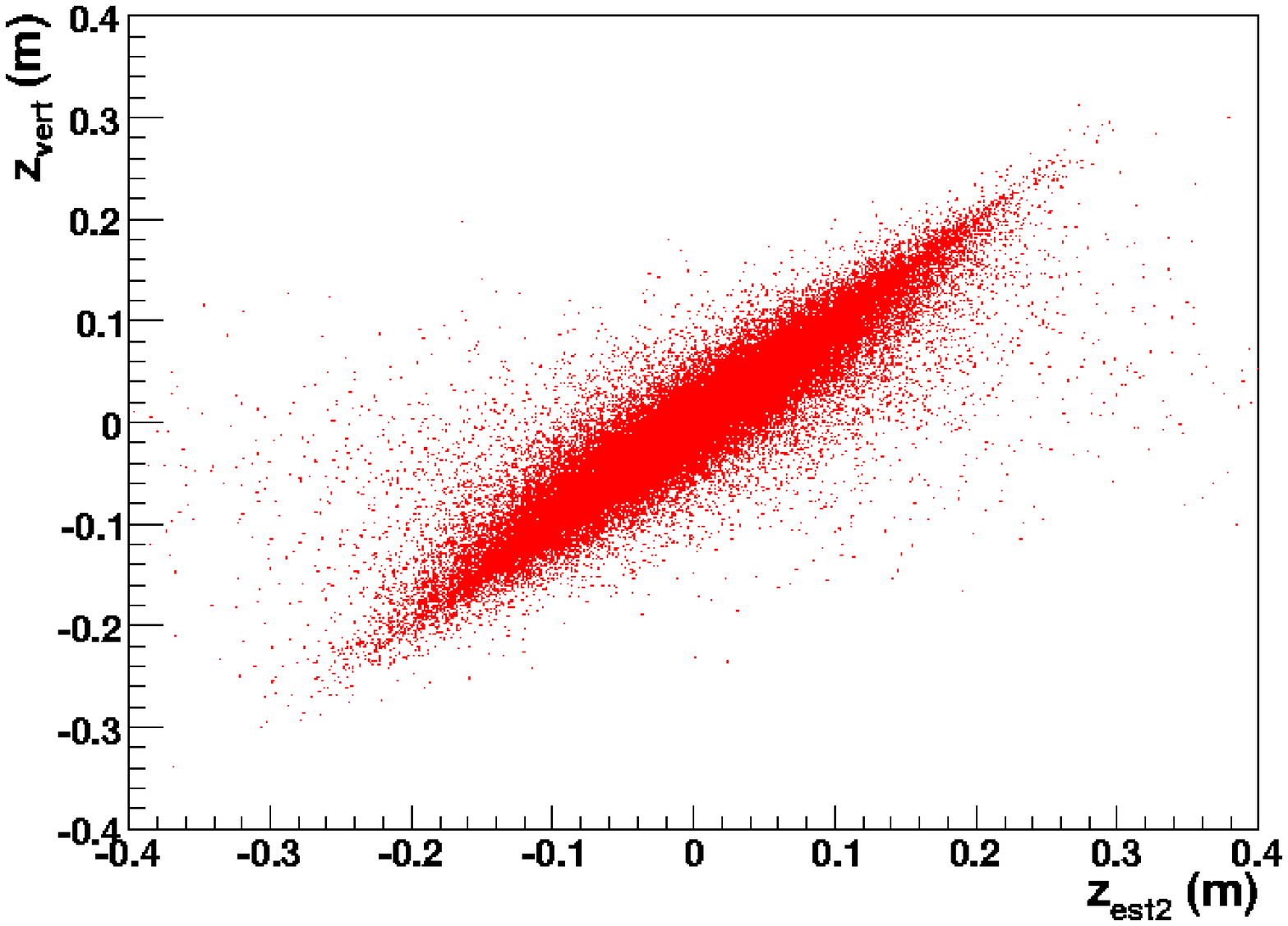, width=80mm,height=70mm}
}}
\put(75, 0){\makebox(0,0)[lb]{
\epsfig{file=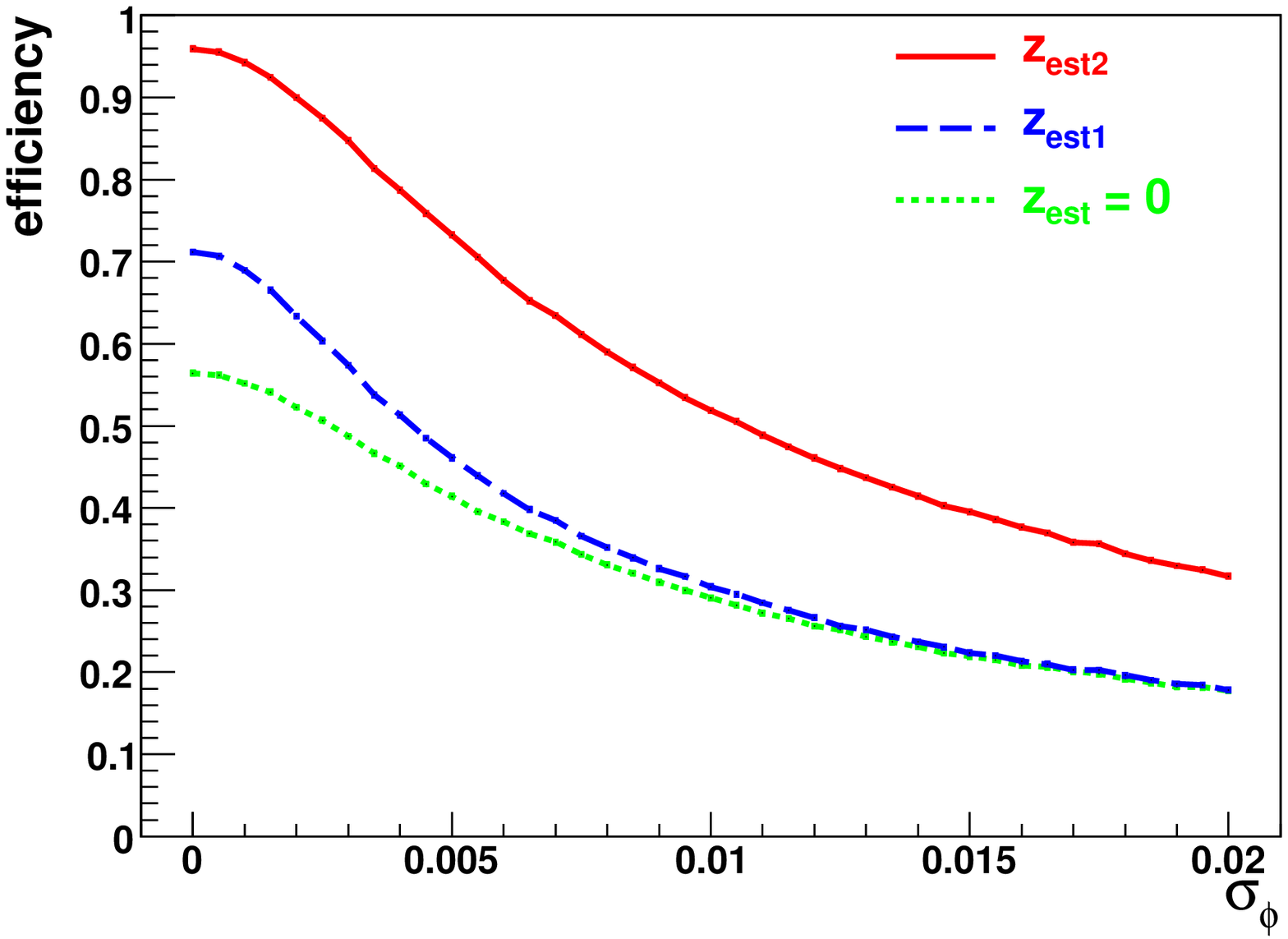, width=80mm,height=70mm}
}}
\put( 40,75){\makebox(0,0)[cb]{\bf (a)}}
\put(115,75){\makebox(0,0)[cb]{\bf (b)}}
\put( 40,-5){\makebox(0,0)[cb]{\bf (c)}}
\put(115,-5){\makebox(0,0)[cb]{\bf (d)}}
\end{picture}
\end{center}
\caption{Studies of reconstruction methods of the reduced pair acoplanarity 
for realistic LHC bunches:
(a) -- the correlation between the generated, $z_{vert}$, and the reconstructed, $z_{est1}$,  vertex positions.
(b) -- the correlation between the generated, $z_{vert}$, and the reconstructed,, $z_{est2}$  vertex positions.
(c) -- a comparison of the estimated vertex position resolutions for the above two reconstruction methods,
(d) -- the efficiency, $\epsilon(\sigma_\phi)$, as a function of the detector azimuthal resolution, $\sigma_\phi$, for the three methods of the vertex position reconstruction: $z_{rec}= z_{est0}=0$ (the dotted line), $z_{rec}=z_{est1}$, (the dashed line), $z_{rec}=z_{est2}$, (the solid line).} 
\label{plot9}
\end{figure}

Figures \ref{plot9}a and \ref{plot9}b present the correlation between the real 
and the reconstructed $z$ position of the vertex using the
two reconstruction methods, respectively.  In Fig.~\ref{plot9}c the 
projections of these scatter plots are compared. The reconstruction precision 
of the vertex $z$ position determines the reconstruction precision of 
the azimuthal angles of the particles hence 
the initial pair $\delta\phi_r^{rec}$. In Fig. \ref{plot9}d the detector resolution dependent 
efficiency estimators, $\epsilon(\sigma_\phi)$, are compared for the three methods of the vertex z position 
reconstruction. If $z_{est2}$ is used,  approximately 90\% signal pair selection efficiency can be achieved 
provided that the hits azimuthal position  is measured with the accuracy better than 2~mrad. If no hit timing measurement is 
available and the nominal vertex position is used,  the efficiency drops by a factor of about two for the same hit  position resolution. 
The event-by-event  reconstruction of the vertex 
position using the detector timing will be particularly important  if the 
longitudinal emittance of the LHC beams will be worse than that assumed in present 
studies.

\section{The dead material effects}
\label{sec:dead}

So far an ideal case of particles propagating in vacuum 
was considered. In reality 
particles traverse  the beam pipe and all the elements 
of the host detector trackers before reaching 
the luminosity detector fiducial volume.
The effects of multiple scattering and radiation 
in the corresponding dead material are evaluated in this section. 

The studies  of the importance of the dead material effects were made 
for the following three values of the dead material thickness
expressed in radiation length units, $X_0$: 
$X/X_0 = 0.2, 0.5, 0.9$. This spread is covering the dead material budget 
of the ATLAS detector in the relevant $\eta$ range.  The dead material 
was assumed to be uniformly distributed along the particle
trajectory\footnote{This rather crude approximation of the reality 
is sufficiently precise  for the studies presented in this paper. In the 
real experimental analysis the distribution of the dead material 
will be determined precisely using the gamma ray conversion tomography
of the host detector tracker.}.
The luminosity  detector $z$ planes were assumed to be $0.1\, X_0$ thick.

\subsection{Multiple scattering}

\begin{figure}[t]
\begin{center}
\setlength{\unitlength}{1mm}
\begin{picture}(160,70)
\put(0,0){\makebox(0,0)[lb]{
\epsfig{file=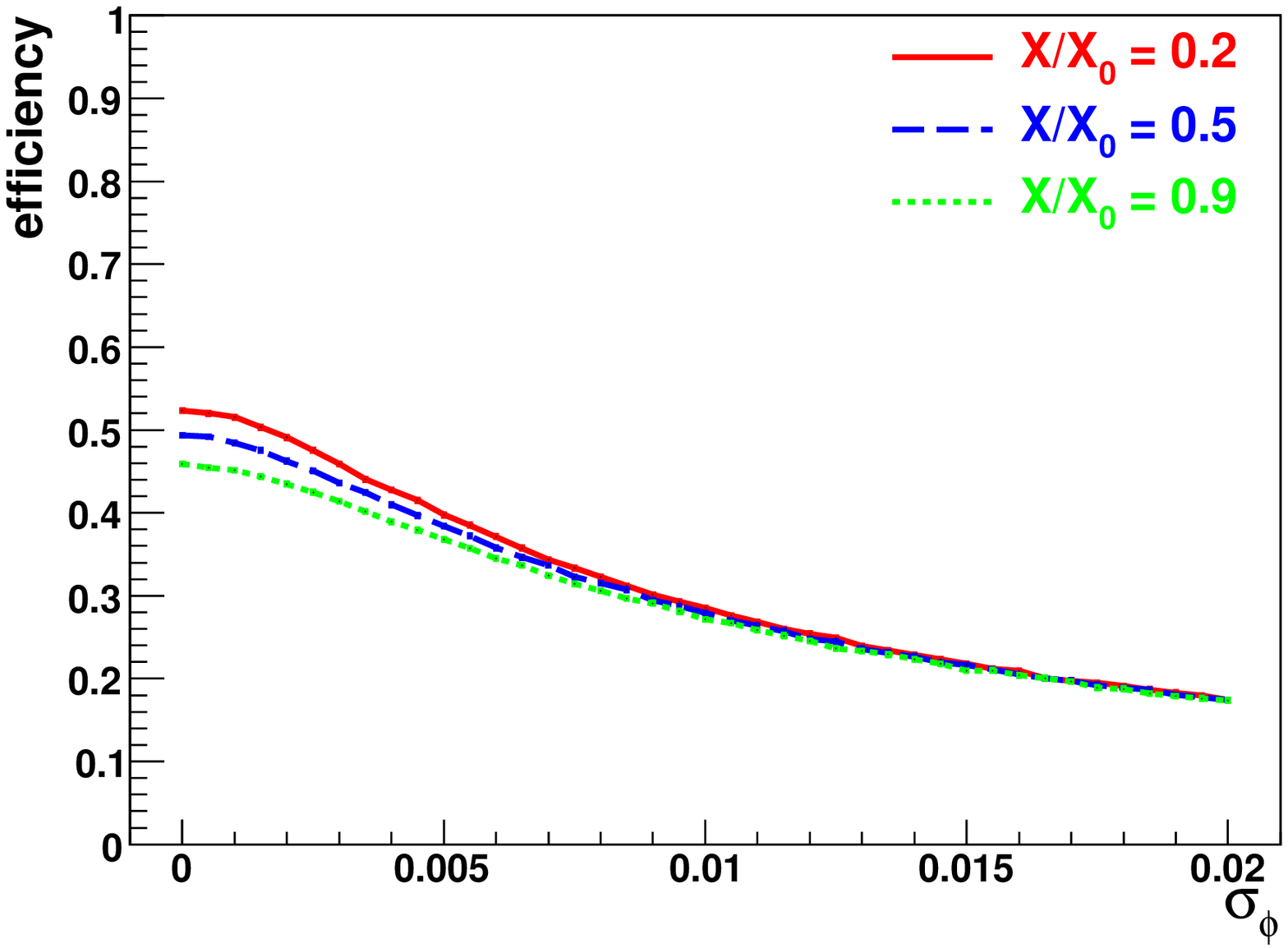, width=80mm,height=70mm}
}}
\put(75,0){\makebox(0,0)[lb]{
\epsfig{file=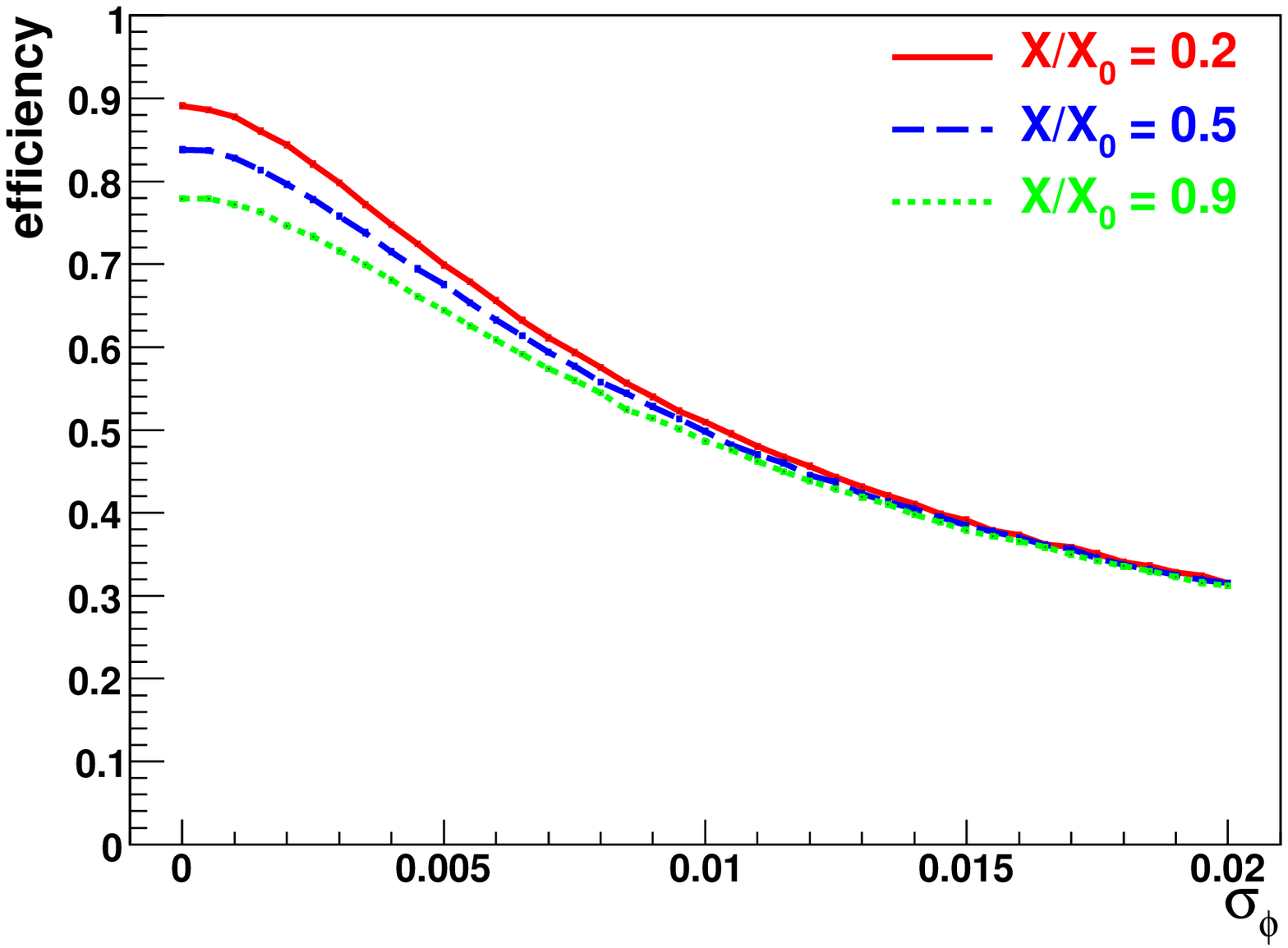, width=80mm,height=70mm}
}}
\put( 40,-5){\makebox(0,0)[cb]{\bf (a)}}
\put(115,-5){\makebox(0,0)[cb]{\bf (b)}}
\end{picture}
\end{center}
\caption{Coulomb multiple scattering studies (LPAIR signal events):
 (a) -- the coplanar pair selection efficiency as a function of the detector azimuthal resolution for the 
three dead material thicknesses expressed in radiation length units, of:  0.2 (the solid line), 0.5 (the dashed line), 0.9 (the dotted line) 
for the reconstructed  vertex position, $z_{rec} = 0$,
 (b) -- as before,  but for the reconstructed  position of the interaction vertex position, $z_{rec}= z_{est2}$.}
\label{plot10}
\end{figure}

The Coulomb multiple scattering distorts the charged particle trajectory.  Hence, 
the vertex-extrapolated  azimuthal angles of the particles, reconstructed using  the luminosity 
detector track segments may be biased. In Figures 
\ref{plot10}a and \ref{plot10}b the coplanar pair selection efficiency is plotted for the 
three radiator thicknesses of 0.2, 0.5 and 0.9 of $X_0$ as a function of the 
detector azimuthal angle measurement resolution. In Fig. \ref{plot10}a the 
efficiency was calculated using the $z_{rec}= 0$ vertex position.
This plot demonstrates that multiple 
scattering does not influence strongly the efficiency dependence 
upon the detector hit azimuthal resolution. Figure \ref{plot10}b 
shows the corresponding efficiencies  using the reconstructed,  $z_{rec}=z_{est2}$, 
position of the interaction vertex.  For the maximal dead material budget the efficiency reaches 80\%, in the case of a  
perfect detector and drops to about 30\% for the hit resolution of $\sigma_\phi = 0.02$ radians. 

\begin{figure}[t]
\begin{center}
\setlength{\unitlength}{1mm}
\begin{picture}(160,70)
\put(0,0){\makebox(0,0)[lb]{
\epsfig{file=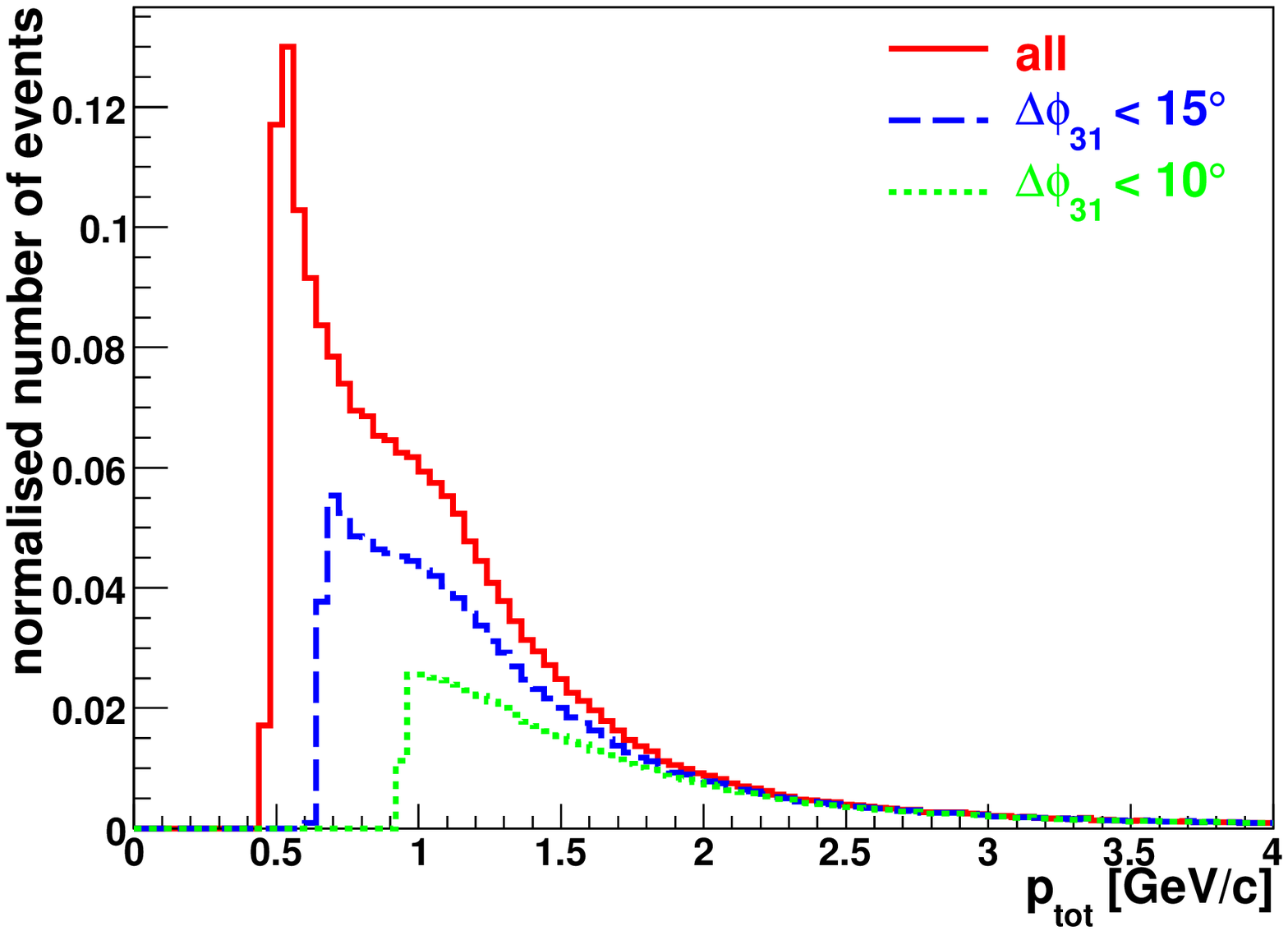, width=80mm,height=70mm}
}}
\put(75,0){\makebox(0,0)[lb]{
\epsfig{file=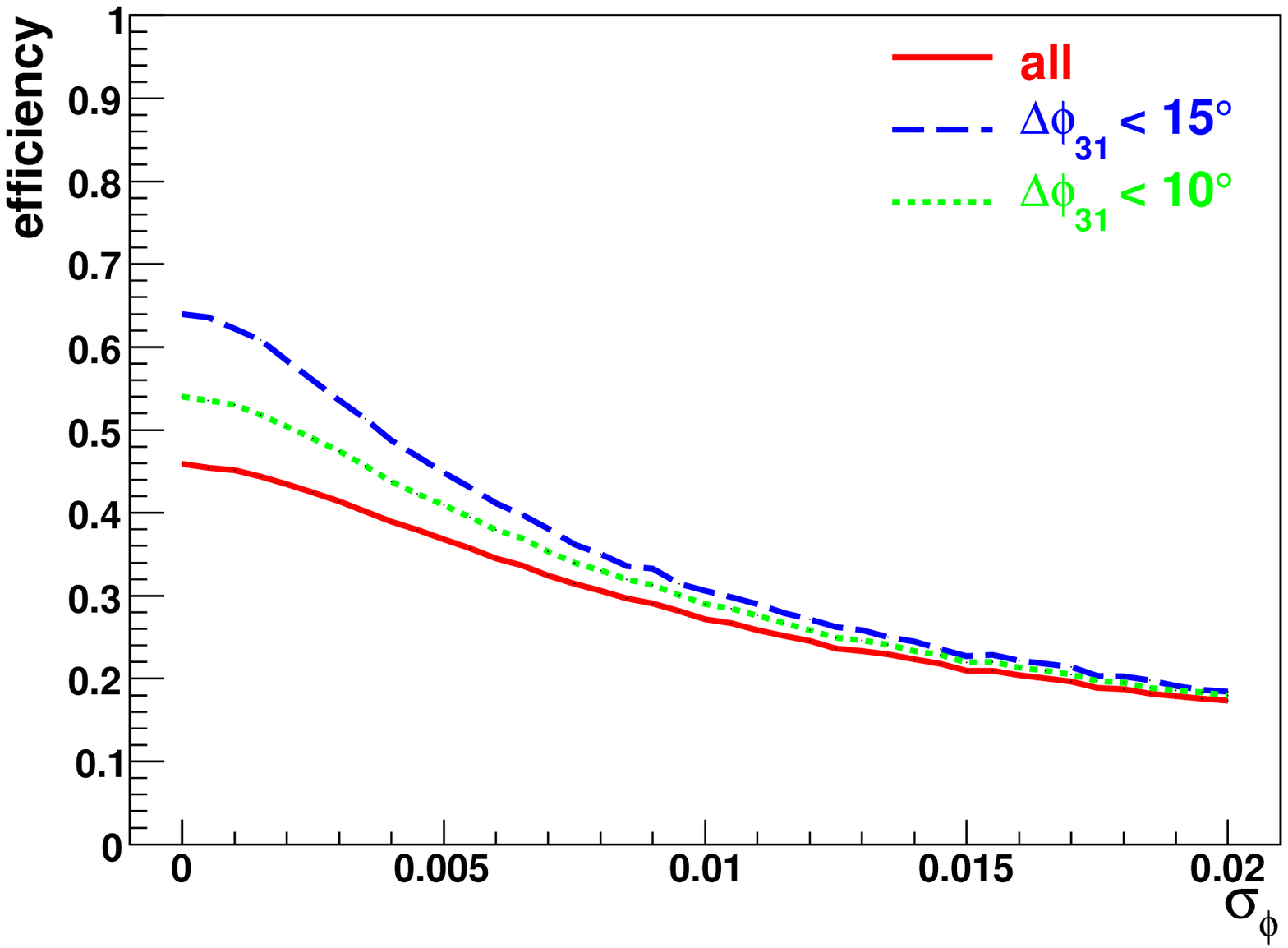, width=80mm,height=70mm}
}}
\put( 40,-5){\makebox(0,0)[cb]{\bf (a)}}
\put(115,-5){\makebox(0,0)[cb]{\bf (b)}}
\end{picture}
\end{center}
\caption{The effective particle momentum cut-off and its effect on
the efficiency of the coplanar pair selection in the presence 
of the Coulomb multiple scattering effects (LPAIR signal events):
(a) -- the particle momentum distribution for the full sample tracks 
(the solid line) and for a subsample of tracks satisfying the following two constraints: $\Delta\phi_{31} < 10^o$  
(the dotted line) and $\Delta\phi_{31} < 15^o$ (the dashed line),
(b)--  efficiency, $\epsilon(\sigma_\phi)$,  as a function of the
detector azimuthal resolution for the three values of
$\Delta\phi_{31}$ cut: no cut (the full line), $\Delta\phi_{31} <
10^o$ (the dotted line) $\Delta\phi_{31} < 15^o$ (the dashed line),
the $z_{rec}=0$ was assumed.}
\label{plot11}
\end{figure}

Since the Coulomb multiple scattering 
distorts predominantly the low momentum particle tracks 
its effects can be reduced further by restricting our sample to   
high momentum particles. A luminosity detector estimator of the particle momentum,  
which can be directly used by the luminosity detector LVL1 trigger logic,  
is the  angular distance of particle hits in the third and 
in the first detector plane, 
$\Delta\phi_{31}^i = \phi_3^i-\phi_1^i$.  Its correlation with particle 
momentum is illustrated in Fig. \ref{plot11}a where 
the distribution of the particle momentum distribution is shown for the three 
different cuts on $\Delta\phi_{31}$: $no~cut$, $\Delta\phi_{31} < 15^\circ$ and 
$\Delta\phi_{31} < 10^\circ$.  In the case of 
$\Delta\phi_{31} < 15^\circ$ only particles of momenta $p_{tot} > 
0.6$~GeV/c are accepted, while the requirement of 
$\Delta\phi_{31} < 10^\circ$ cut-off moves this limit to almost 1~GeV/c.

The efficiencies calculated for the above three subsamples of events, for $z_{rec}=0$, 
 are compared in Fig. \ref{plot11}b. As can be 
observed the efficiency increases with decreasing value of the cut on 
$\Delta\phi_{31}$ and for a perfect detector it changes between 45\% and 65\% 
for the $\Delta\phi_{31}$ cut-off value ranging from $no-cut$ to $10^\circ$. 
The efficiency decreases with deteriorating azimuthal resolution and 
reaches about 20\% for $\sigma_\phi = 0.02$ radians irrespectively of the 
$\Delta\phi_{31}$ cut value.
The effective particle momentum cut reduces not only  
the impact of the multiple scattering effects on the coplanar pair selection 
efficiency. It reduces as well  the sensitivity of the selection efficiency to the 
reconstruction precision of the collision vertex position.

\subsection{Radiation}

The effects of multiple scattering concerns both the $e^+e^-$ and the $\mu^+\mu^-$ pairs. In this 
section the radiation effects are evaluated. These effects are of importance only for the $e^+e^-$ pairs. 

The  radiation of photons leads to the electron energy losses in the dead 
material. Overwhelming majority of photons are emitted co-linearly to the electron trajectory.
In the absence of magnetic field the electron azimuthal angle is unchanged by the photon emission
and the reconstructed pair acoplanarity remains unaffected. 

In the presence of magnetic field this is no longer the case, in particular for catastrophic 
losses of the electron energy associated with radiation of hard photons. 

The radiation gives a rise to the following  two effects. Firstly,  the electrons
produced in the acceptance region of the luminosity detector may no longer 
reach its  fiducial volume. Secondly, even, if they reach its  fiducial volume 
their  angular velocities, $\omega(B, p_T)$ will be different from the initial ones. 
Consequently, the 
extrapolated pair acoplanarity could deviate  significantly from the true, initial 
one. Both effects reduce the coplanar pair selection efficiency. 

\begin{figure}[ht]
\begin{center}
\setlength{\unitlength}{1mm}
\begin{picture}(160,70)
\put(0,0){\makebox(0,0)[lb]{
\epsfig{file=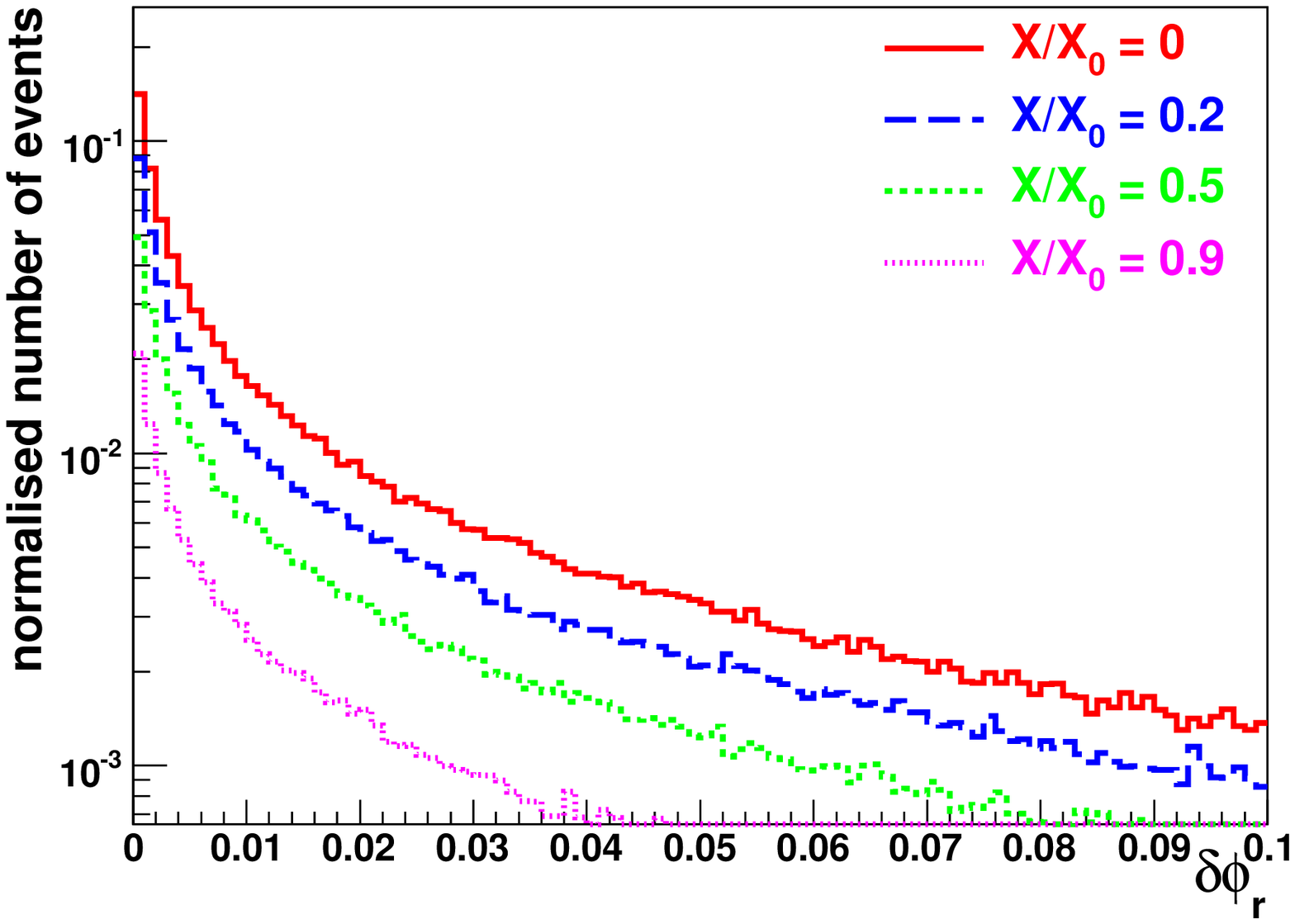, width=80mm,height=70mm}
}}
\put(75,0){\makebox(0,0)[lb]{
\epsfig{file=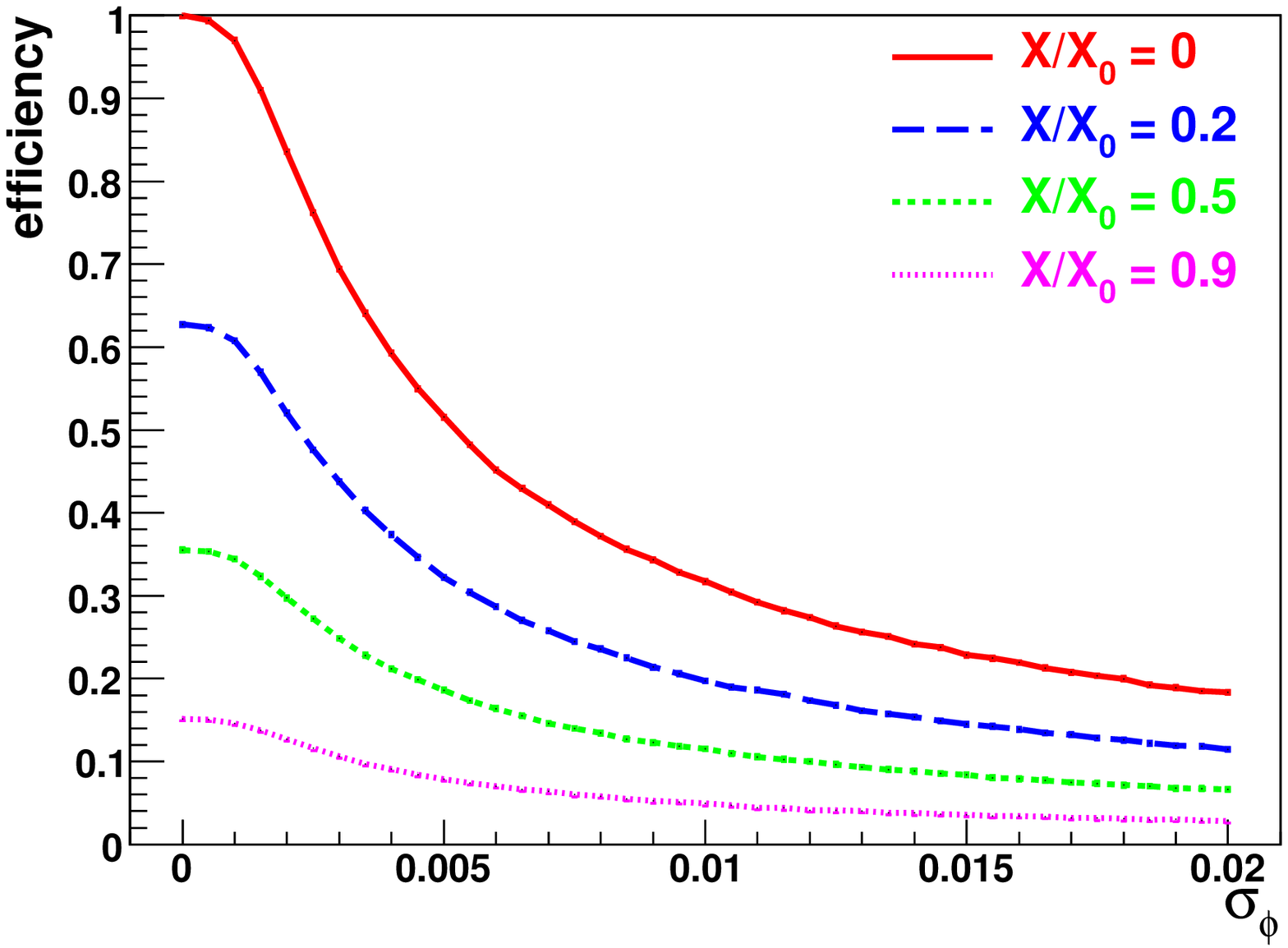, width=80mm,height=70mm}
}}
\put( 40,-5){\makebox(0,0)[cb]{\bf (a)}}
\put(115,-5){\makebox(0,0)[cb]{\bf (b)}}
\end{picture}
\end{center}
\caption{The study of the radiation effects on the coplanar lepton pair selection efficiency: 
(a) -- the reduced acoplanarity distribution at the interaction vertex for the four thicknesses of the dead
 material in units of $X_0$:
0 (the solid line), 0.2 (the dashed line), 0.5 (the dotted line), 0.9 
(the dash-dotted line),
(b) -- the  coplanar pair efficiency, $\epsilon(\sigma_\phi)$ as a function of the detector 
azimuthal resolution,  for the above four thicknesses of the dead material.}
\label{plot12}
\end{figure}

In Fig. \ref{plot12}a the distributions of the initial reduced acoplanarity of 
the electron pairs are shown for four different radiator thicknesses. 
These distributions are sensitive to the dead material distribution on the 
path of particles from the production vertex up to the luminosity detector fiducial volume.
The largest impact is due to the  dead material in the vicinity of the collision vertex,
because it influences a large fraction of the particle path. 
The distributions shown in  Fig. \ref{plot12}a were obtained assuming that the radiator is placed
at  the collision vertex\footnote{Precise simulations including the realistic dead material distribution are beyond
the scope of this paper.}. They represent,  thus,  the maximal losses of the signal pairs. 
 
Figure \ref{plot12}b shows the efficiencies, $\epsilon(\sigma_\phi)$, as a 
function of the detector azimuthal resolution for the four values of the 
radiator thickness.  The function drawn with the solid line represents the 
ultimate efficiency function for the zero thickness of the radiator.  The 
radiator presence leads to lower efficiency values. In the case of $0.9\,X_0$ thick 
radiator the perfect detector efficiency drops to about 15\% and its value is 
reduced  to 5\% for $\sigma_\phi = 0.02$ radians.

\subsection{Multiple scattering and  radiation}

\begin{figure}[ht]
\begin{center}
\setlength{\unitlength}{1mm}
\begin{picture}(160,70)
\put(0,0){\makebox(0,0)[lb]{
\epsfig{file=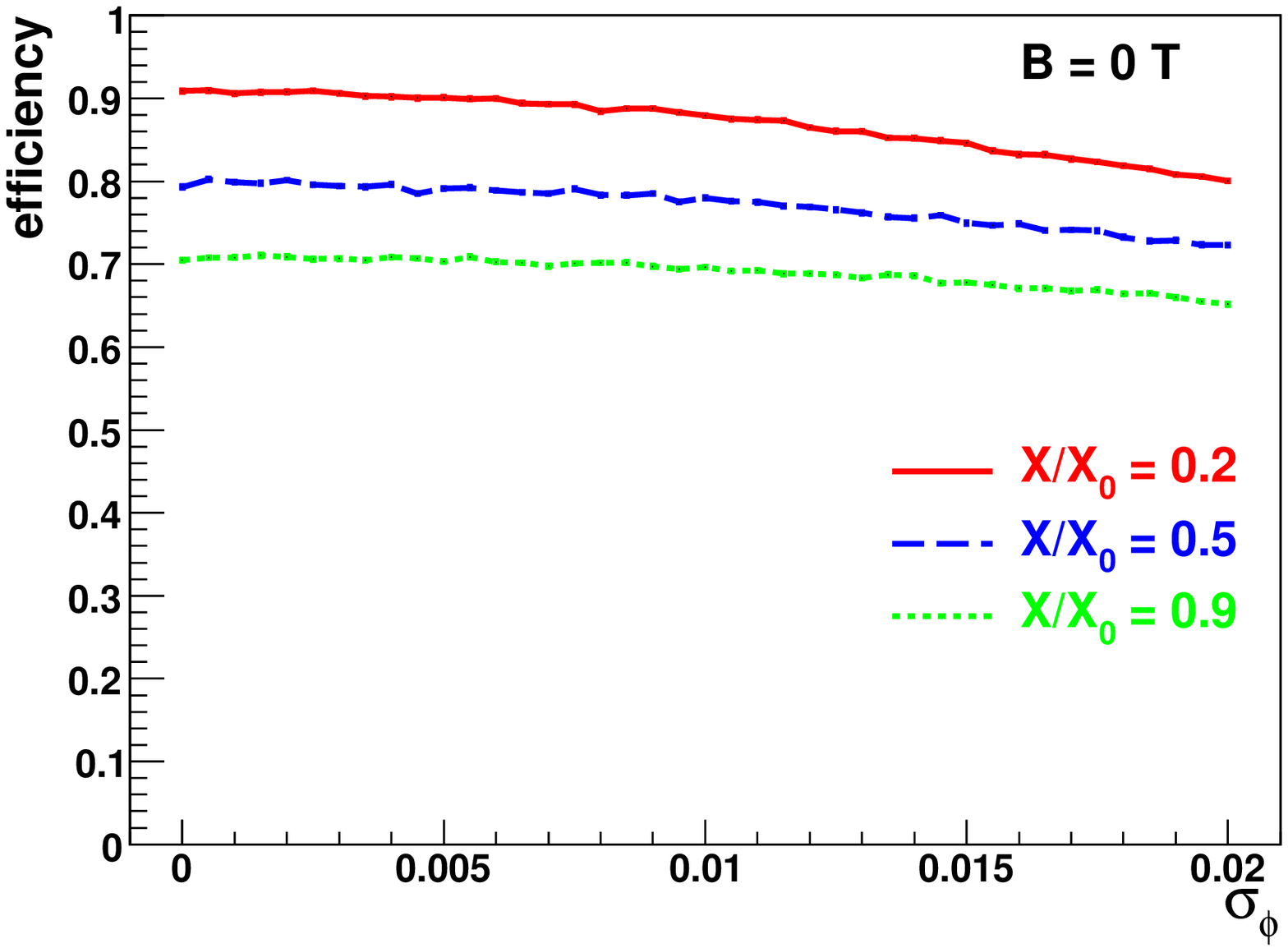,width=80mm,height=70mm}
}}

\put(75,0){\makebox(0,0)[lb]{
\epsfig{file=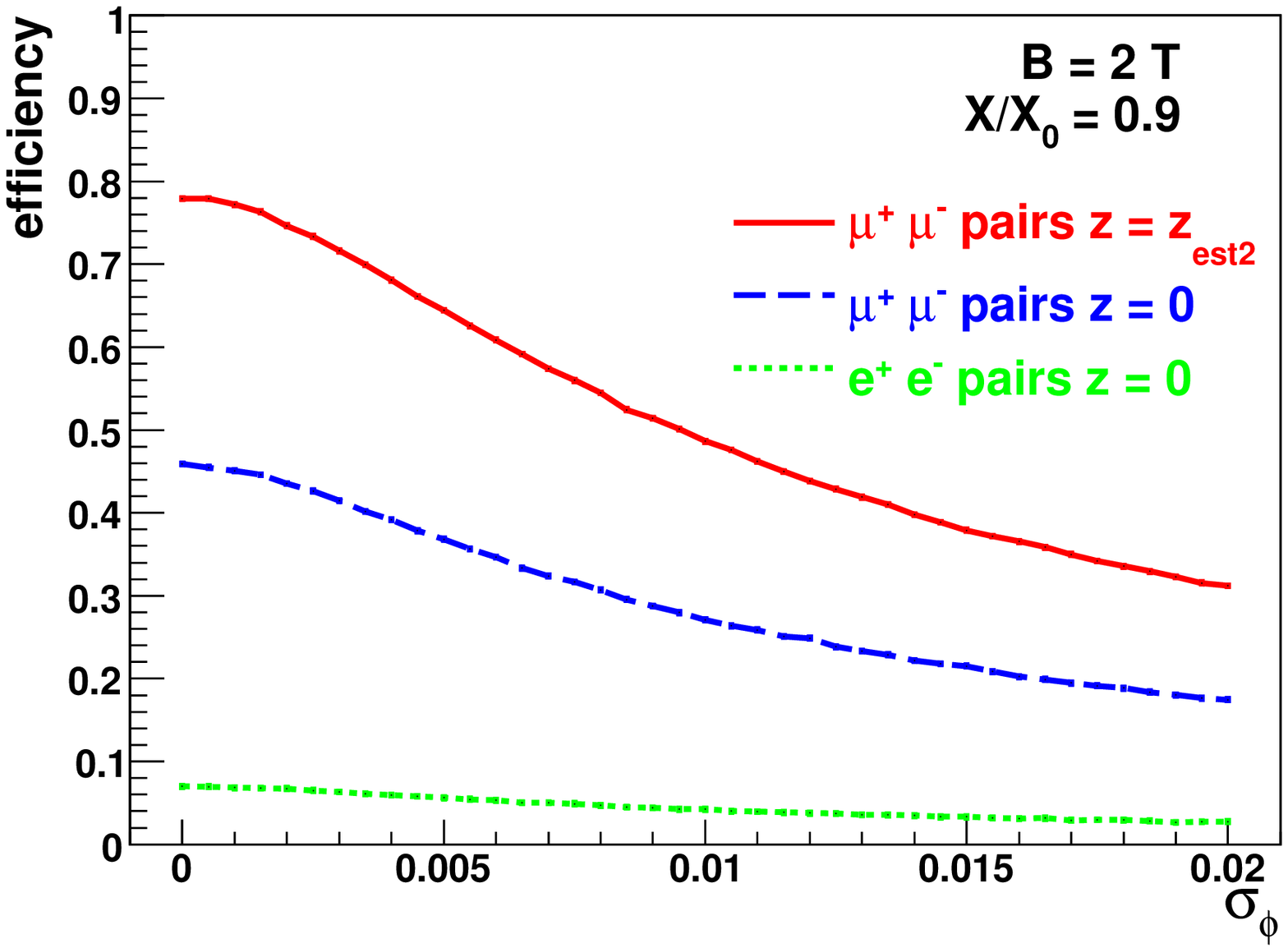,width=80mm,height=70mm}
}}
\put( 40,-5){\makebox(0,0)[cb]{\bf (a)}}
\put(115,-5){\makebox(0,0)[cb]{\bf (b)}}
\end{picture}
\end{center}
\caption{The efficiency, $\epsilon(\sigma_\phi)$, as a function of the detector azimuthal resolution, $\sigma_\phi$:
(a) -- for the $B0$ field configuration, and for the three values of the dead material thickness expressed 
in the radiation length units: 0.2, 0.5 and 0.9 (the dashed, the dotted and the dash-dotted lines, respectively),  
(b) -- for the {\bf B2} field configuration and for the dead material thickness of 0.9. The solid and the dashed lines, 
for the opposite charge muon pairs,  correspond to, respectively,  the $z_{rec} = z_{est2}$ and $z_{rec} = 0$ values.
The dotted line, for the electron-positron pairs,  correspond to  $z_{rec} = 0$ value.}
\label{plot13}
\end{figure}

The coplanar pair selection efficiencies, $\epsilon(\sigma_\phi)$, plotted as a function of the detector 
azimuthal resolution,  for the  three values of the dead material  thickness,  and for the {\bf B0} 
field configuration,  are  presented in Fig. \ref{plot13}a. The pair acoplanarity 
was reconstructed  from  the average azimuthal positions of the hits left by the 
leptons in the detector layers. The efficiency is high and depends  weakly 
on the detector azimuthal resolution. If the radiator is $0.9\,X_0$ thick the 
efficiency is about 70\% and practically does not depend on the detector $\phi$ 
measurement resolution.  This figure demonstrates that in absence of the 
detector magnetic field:
\begin{itemize}
\item the extrapolation of tagged tracks to a common vertex is not indispensable 
to determine their initial acoplanarity,
\item the measurements are largely insensitive to the photon radiation effects,
\item the effects of multiple scattering is more pronounced than in the case
of the {\bf B2} field configuration (small momentum particles are no longer 
swept out by the magnetic field) 
\item the coplanar lepton pairs can be efficiently selected by the luminosity 
detector even if its azimuthal angle measurement resolution is very modest.
\end{itemize}

The presented efficiencies are the same for the  $e^+e^-$ and the $\mu^+\mu^-$ pairs. 
This is no longer the case for the {\bf B2} field configuration. 

The efficiencies modified by the multiple scattering and radiation effects, for the {\bf B2} field configuration are presented  in 
Figure \ref{plot13}b.  Again, the efficiency, $\epsilon(\sigma_\phi)$, is shown as a function of the detector 
azimuthal angle measurement resolution.  All distributions correspond to  the maximal dead material budget, $X/X_o = 0.9$, 
and to the worst case of the dead material concentrated in the vicinity of the interaction vertex.

The solid line represents to the coplanar pair selection efficiency for the $\mu^+\mu^-$ pairs
for the luminosity detector capable to measure  both the azimuthal position 
of the tagged particle hits and their timing. We recall that these measurements allow for a precise 
reconstruction of the position of the interaction vertex:  $z_{rec} = z_{est2}$. 
The efficiency reaches 80\%  and drops to about 55\% for $\sigma_\phi = 0.02$.
The dashed line shows the efficiency for the $\mu^+\mu^-$pairs
for a detector which does not provide the timing measurement of the particle hits. 
In this case,  the event-by-event reconstruction of the vertex position cannot be made and 
its nominal position,  $z_{rec} = 0$,  is assumed for each of the observed pairs. 
The efficiency reaches 46\% for an ultimate precision detector and drops to about 20\% for 
$\sigma_\phi = 0.02$.
The dotted line shows the efficiency for the $e^+e^-$pairs 
for a detector which does not provide the timing measurement of the tagged particle hits. Again, the 
nominal position $z_{rec} = 0$ value for each of the observed pairs is assumed.
The drop of efficiency is driven basically  by the radiative processes.
The efficiency reaches 8\% for an ultimate precision detector and drops to about 4\% for 
$\sigma_\phi = 0.02$

Studies presented in this section points out to the three emerging methods  of the luminosity 
measurement. Each of them puts different emphasis on different  performance aspects of 
the luminosity detector.  For the measurement using muon pairs a  precise hit timing 
measurement and a  high azimuthal granularity of the luminosity detector are crucial  to achieve 
the highest possible selection efficiency of coplanar pair selection. This is important 
because only small fraction of large momentum muons will be able reach the muon spectrometer and 
be subsequently identified and selected 
by the LVL2 and the EF triggers of the host detector\footnote{Note that the position of the 
luminosity detector at the $\eta$-extremity of the host detector tracker maximises muon total momentum 
for a fixed transverse momentum value.}. 
For the measurement using  $e^+e^-$ pairs the detector timing functions and its fine $\phi$-granularity 
are less  important because the coplanar pair selection efficiency is driven mainly by the  
electron radiation in the dead material of the host detector. 
The same conclusion can be made for the luminosity measurement in  the {\bf B0}-periods of the host detector operation,
even if for totally different reasons. It is important to note that  
the high  coplanar pair selection efficiency in the {\bf B0} periods allows to achieve a 
comparable statistical precision of the luminosity measurement in about ten times shorter 
time intervals, than in the case of the {\bf B2} periods.

\section{The trigger and the data acquisition system requirements}
\label{sec:TDAQ}

\subsection{Operation aspects} 

In previous sections the timing and the space resolution requirements for the
luminosity detector were  discussed. They were driven basically 
by  its capacity to select  coplanar lepton pairs with the highest 
achievable efficiency.  These requirements have to complemented by
the requirements that  the 
coplanar lepton pairs must be selected within the host detector LVL1 latency time 
and that  the selection process must be monitored with adequate precision. 
These above two requirements  will determine the trigger and the data acquisition 
performance requirements discussed in this section. 
 
Ideally, for a noiseless detector and for perfectly collimated beams 
the silent bunch crossings could be identified by the LVL1  trigger logic 
of the luminosity detector by demanding  zero multiplicity of the 
track segments both in the left and in the right arm of the luminosity detector.
Event candidates for the luminosity measurement could be identified 
by the LVL1 trigger logic  as those with  
two  track segments in the left (right)  arm of the luminosity detector and 
no track segment in the other arm. 
This LVL1 sample of preselected events would  be still dominated by the 
"non-silent" bunch crossings, in which the luminosity detector signals are due
to hadrons produced in diffractive strong interaction collisions. However,  
by demanding that the vertex extrapolated  pair 
acoplanarity satisfies the condition: $\delta\phi_r  < 0.01$ the rate of these background 
events can be  reduced, using solely the luminosity detector data, to the level of few kHz in 
a wide range of the machine luminosities. 
This rate, could  be reduced 
further by the Central Trigger Processor (CTP)  of the host detector, by using the full set of the host detector LVL1
trigger bits. The corresponding reduction factor could  be dynamically adjusted to 
the allocated share of the host detector trigger and data acquisition capacities 
such that the LVL1 accepted luminosity events  represent a "non-interfering" 
fraction of the host detector LVL1 accepted events. 

In the LVL1-selected luminosity event candidates sample, the hadron background events 
outnumber the $l^+l^-$ signal events still by a factor of $\cal{O}$($10^5$). Their 
rate could  be reduced to a 
signal event rate of $\cal{O}$(0.1~Hz) using the present capacity of the HLT 
system of the host detector. At such a small rate, recording the full host and 
luminosity detector information by the host detector Data Acquisition (DAC) 
system would hardly interfere with the host detector  canonical operation modes.

 The final sample of the HLT selected luminosity event candidates would be 
 populated  mostly the peripheral electromagnetic collision events in which a coplanar lepton pair is 
 produced. Their final analysis could be done off-line concurrently with the analysis 
 of any user defined subsample of events providing the absolute normalisation of 
 the measured cross section independently of the data quality criteria. 
 
In reality,  the luminosity detector operating conditions may be different from 
the ideal ones and varying with  time, instantaneous bunch-by-bunch luminosity, 
beam currents, $\beta ^*$, the beam crossing angle and with other parameters. 
In addition,  the detector may have the periods of noise producing spurious track segments. 
Therefore, the luminosity detector  LVL1 trigger algorithm must be able to identify 
the coplanar pairs in the bunch crossings where several track segments, 
not associated with beam-beam collisions,  are observed in each of the 
luminosity detector arms.  Since the algorithm processing time increases with the square of the 
number of tack segments this remedy has a sharp processing power dependent limit. 

For a  scheme to be robust against any variation of the  data taking conditions, supplementary
``rate-stabilising" methods are required. The first one, protecting the triggering scheme 
against beam halo track candidates is to validate the track segments with the precise 
time stamps. As discussed earlier this requires the hits in the $z_1$ plane to be  measured
with 0.5 ns resolution. In the periods of large machine noise the search for coplanar 
lepton pairs could be made using track segments with the beam-beam interaction window
time stamp.

 Another protection against periods with large number of spurious track segments 
would be to use in the searches of coplanar pairs only those  of the track segments 
which are likely to be left by the electrons. This remedy would require either 
a dedicated luminosity detector technology or an extension of the present object-multiplicity 
driven LVL1 logic of the host detector CTP to a scheme in which the topological 
association of the trigger elements is made. These aspects are beyond the scope of 
this series of papers and will not be discussed here.

\subsection{Monitoring aspects}

In order to achieve the luminosity measurement precision target the 
event selection process will have to be precisely monitored and the event 
selection efficiency, and acceptances are required to be determined directly from 
the data.  In addition, the background subtraction scheme is required to be  
independent of the modelling precision of the strong interactions of the colliding protons.  
Those criteria drive the principal performance requirements for 
the DAQ system of the luminosity detector and constrain its incorporation 
within the DAQ system of the host detector.   

Fulfilling these requirements if facilitated in the proposed scheme in which  
the luminosity detector  is embedded within the fiducial volume of the tracker
of the host detector. As discussed earlier, owing to such a configuration all 
the particles crossing the luminosity detector will 
have their tracks reconstructed and the energy deposits measured by the host 
detector. This will allow to use a large sample of reconstructed minimum bias 
events,  associated parasitically with any type of selected events,  for the data 
driven background subtraction and for the data-driven determination of the 
event selection efficiencies. Owing to a large statistics of such events, the 
particle identification capacity of the host detector,  and the use of the hadron resonances
as the tagged sources of the all the particle species,  
 the lepton pair selection efficiencies and the hadron background can be precisely controlled on the 
bunch-by-bunch basis in a wide range of the instantaneous  
luminosities. 

The principal requirements are thus confined merely to the performance aspects 
of the DAQ system of the luminosity detector, in particular to its efficiency in  collecting of the 
dedicated samples of the monitoring events. 

For example, it is  required that the number of  validated "in-time" and "out-of-time" track 
segments in the $\phi$-sectors of the luminosity detector is monitored by the  
local data acquisition system to provide a very fast, bunch-by-bunch relative 
luminosity determination. The DAQ system of the luminosity detector must 
provide a precise monitoring of the time evolution of the rate of the silent 
bunch crossing both for each of the interacting and the pilot (those which do 
not have bunch partners to collide) bunches. To achieve this goal the silent 
bunch crossings will have to be  monitored locally, and controlled globally using random events collected by the host detector. 
 
\section{Conclusions and outlook}

In our  previous paper  \cite{first},  the phase-space region of the lepton pair production process
$pp \rightarrow l^+l^- + X$  was selected.  We have shown that its rate can be theoretically controlled at the LHC to 
a $ \leq 1\%$ precision.  
In the present  paper a realistic proposal of the selection strategy of a significant fraction of 
events produced in this region was presented.  

The main obstacle for the present LHC detectors to implement such a scheme is a 
missing capacity of their LVL1 trigger to reduce by seven orders of magnitude the rate 
of the background hadron pairs produced in ordinary strong interaction mediated collisions of the beam particles. 

A remedy proposed in this paper is based on two ideas.  The first one  is to select already by the LVL1 trigger only events with back-to-back pairs,  leaving the lepton identification to the subsequent event selection stages. The second one is to 
drastically reduce the number of bunch crossings in which coplanar pairs are searched to the 
subsample of "silent bunch crossings" and monitor their frequency using the random bunch-crossing trigger.  

Their implementation requires an upgrade of one of the LHC detectors by incorporating 
within its fiducial volume a dedicated "luminosity detector". Its main goal is to analyse,  
within the LVL1 trigger latency,   the topology and the origin of the particle hits
detected in its fiducial volume, and to deliver  the LVL1 accept/reject  signal to the CTP unit of the host detector. 

The minimal requirements for the timing and for the spacial hits- resolutions were analysed in a realistic operation environment, including a finite-size particle bunches and 
the effects due to dead material. It was demonstrated that for a very modest 
detector resolution  requirements the lepton pair candidate events 
can be efficiently selected. 

An  implementation of the proposed scheme based on a concrete model 
of the luminosity detector LVL1 trigger and including a  complete, host detector signal based,  
LVL2 trigger and Event Filter  selection procedure of the 
luminosity events will be  presented in the forthcoming  paper.


\begin{thebibliography}{999}

\bibitem{first} M. W. Krasny, J. Chwastowski and K. S{\l}owikowski,
Nucl. Instrum. Meth. {\bf A584}~(2008)~42. 
%
\bibitem{ATLAS} The ATLAS Collab., G. Aad et al., 
J. Inst. {\bf 3} (2008) S08003.
%
%
\bibitem{ATLASTDAQ} ATLAS Collab.
CERN-LHCC-2003-022.
%
\bibitem{LPAIR} S. P. Baranov, O. Dunger, H. Shooshtari and 
                J. A. M. Vermaseren, 
                Proc. of Physics at HERA, vol. 3, (1992) 1478.
%
\bibitem{PYTHIA}  T. Sj\"ostrand, P. Ed\'en, C. Friberg, L. L\"onnblad, 
                G. Miu, S. Mrenna and E. Norrbin, Computer
                Phys. Commun. {\bf 135} (2001) 238.
%
\bibitem{Geant} R. Brun et al., Geant 3.21, CERN Program Library Long Writeup W5013,\\
Geant4 Collab., S. Agostinelli et al., Nucl. Instrum. Meth. {\bf A506} (2003) 250,\\
Geant4 Collab., J. Allison et al., IEEE Trans. Nucl. Science {\bf 53} (2006) 270.
%
\bibitem{LHC} LHC White Book, CERN/AC/93-03,\\
LHC Conceptual Design Report, CERN/AC/96-05.
%
\bibitem{PDG} Particle Data Group, C. Amsler et al., Phys. Lett. {\bf B667} (2008) 1.
%
\bibitem{tsai} Yung-Su Tsai,
Phys. Rev. {\bf 149} (1966) 1248,\\
Rev. Mod. Phys. {\bf 46} (1974) 815; erratum Rev. Mod. Phys. {\bf 49} (1977) 421.
%
\end{thebibliography}
\end{document}